\documentstyle[]{mn}

\input{psfig.sty}

\ifnfsstwo

\fi

\ifnfssone
  \newmathalphabet{\mathit}
    \addtoversion{normal}{\mathit}{cmr}{m}{it}
    \addtoversion{bold}{\mathit}{cmr}{bx}{it}

\fi

\ifoldfss

\fi

\loadboldmathitalic
\loadboldgreek

\def\Zsun{\thinspace\hbox{$\hbox{Z}_{\odot}$}}
\def\msun{\thinspace\hbox{$\hbox{M}_{\odot}$}}

\def\kms{\,km~s$^{-1}$}
\def\cc{cm$^{-2}$}

\def\gal{galaxy}

\def\ev{evolution}
\def\chev{chemical evolution}
\def\chdyn{chemodynamical}
\def\for{formation}
\def\sfor{star formation}
\def\gfor{galaxy formation}

\def\gw{galactic wind}

\title[Nature of DLAs and chemical evolution]
       {Clues to the nature of damped Lyman alpha systems
from chemical evolution models}
\author[G. A. Lanfranchi and A. C. S. Fria\c ca]
 {Gustavo A. Lanfranchi $^1$ and Amancio C. S. Fria\c ca$^1$\\
 $^1$IAG-USP,
 R. do Mat\~ao 1226, Cidade Universit\'aria, 05508-900 S\~ao Paulo, SP, Brazil\\}

\pubyear{2003}

\begin{document}
\maketitle

\begin{abstract}

The evolution of the metallicity
of damped Lyman $\alpha$ systems (DLAs) is investigated
in order to explore several scenarios for the nature of these systems.
The observational data on chemical abundances of DLAs
are analysed with robust statistical methods,
and the abundances are corrected for dust depletion.
The results of this analysis
are compared to predictions of several classes of chemical evolution
models describing a variety of scenarios for DLAs:
one-zone dwarf galaxy models, multi-zone disk models,
and chemodynamical models representing dwarf galaxies.
In order to settle constraints for star formation time scales
and metal production in DLAs, we compare the observational
data on the [$\alpha$/Fe] and [N/$\alpha$] ratios
to the predictions from the models.
In DLAs, these ratios are only partially reproduced by
the dwarf galaxy one-zone model and by the disk model.
On the other hand, the chemodynamical model for dwarf galaxies reproduces
the properties of nearly all DLAs.
The connection between the gas flow evolution and the star formation rate
is the reason for the ability of this model in reproducing
the range of abundance ratios seen in DLAs.
The comparison of the observed [$\alpha$/Fe] and [N/$\alpha$] trends
to the predictions of the chemodynamical model
is used to derive the formation epoch of dwarf galaxies.
The \chdyn\ model predicts that dwarf galaxies make a significant contribution
to the observed total neutral gas density in DLAs,
and that this contribution is more important at high redshifts ($z\ga 2-3$).
This is consistent with a scenario in which the DLA population is dominated 
by dwarf galaxies at high redshifts and by disks at lower redshifts.
The relation between DLAs and Lyman Break Galaxies
(LBGs) is investigated with \chdyn\ models describing LBGs.
Our results calls for a smoother
progression in the evolutionary history of DLAs and LBGs
rather than a sharp dichotomy between the two populations.
LBGs and DLAs may constitute a sequence of increasing star formation
rate, with the LBGs being systems 
with typically short star formation time scales ($\sim10^8$ yr),
and the DLAs having slower star formation.
We also arise the possibility that we could be missing a whole population of
high HI density column objects, with metallicities intermediate
between those of DLAs and LBGs.
Finally, we discuss the possibility 
that relying only on the observations of DLAs
could lead to an underestimate of the metal content of the high redshift Universe.

\end{abstract}


\section{Introduction}

The study of chemical abundances in the QSO absorption line systems (QALs)
and their evolution with redshift are key factors to be 
considered in our attempts to track the history of galaxy formation 
in the Universe. These systems are also essential in tracing
the cosmic chemical evolution because they sample a variety of 
environments and are representative systems of the early Universe,
being enriched by the first stellar generations
(Pettini et al. 1997a, 2000a).

The Damped Lyman $\alpha$ Systems (DLAs) are the QALs with the highest
neutral hydrogen column densities (N(HI) $\ga 10^{20}$ \cc).
These systems are considered to be the most suitable 
for chemical abundance studies in the high-redshift Universe
because they are believed to be the progenitors of present day galaxies,
for a number of reasons.
Owing to their very large N(HI) values the DLAs dominate the neutral
content of the universe (Wolfe et al. 1995; Rao \& Turnshek 2000).
Furthermore, the comoving baryonic mass density in gas inferred from the
DLAs at $z=2-3$ nearly coincides with the current baryonic mass density 
in stars today (Wolfe et al. 1995; Storrie-Lombardi \& Wolfe 2000).

The high density column of these systems allowed abundance
measurements of a number of elements never detected before in
QALs, as is the case of zinc,
which has an important role in DLA studies
(on the other hand, strong lines of abundant elements,
e.g. oxygen, become saturated, making it difficult to measure their abundances).
In addition, their large N(HI) values require 
little ionization corrections (Viegas 1995, Vladilo et al. 2001).
In the last few years, accurate abundances of several chemical elements
have been derived for DLAs at high {\it z} (Lu et al. 1996, Prochaska 
\& Wolfe 1999, Prochaska et al. 2001, Centuri\'on et al. 2000, 
Molaro et al 2001), allowing detailed studies
of chemical evolution in these systems, even though there are uncertainties that
must be taken into account, such as depletion of some 
refractory elements onto dust grains and selection effects.

A widespread indicator of the metallicity in DLAs is the
Zn abundance ([Zn/H]), which is believed to be very little or not
affected at all by dust depletion. The Zn abundances range 
from -2 to 0 (Vladilo 1998, Lauroesch et al. 1996, Pettini et al. 1994, 1997b),
suggesting that DLAs have a range of formation epochs or star 
formation histories. 
The form of the evolution
of metallicity with redshift is not yet clearly known.
Several studies argue that at $z < 2$ the mean metallicity 
of damped systems is approximately  1/10 - 1/30 solar with a large 
scatter and that it decreases at $z \approx 3$, even though the picture 
here is less certain (Pettini et al. 1994, Pettini et al 1997b, 
Pettini 1999, Lu et al. 1996, Prochaska $\&$ Wolfe 1999). By contrast, 
in a more extended work, Prochaska \& Wolfe (2002) concluded that 
there is no significant evolution in metallicity of the DLAs over the
$z \approx 1.5-3.5$ range, although there appears to be
a marked decline at $z>3.5$, but only by a factor of a few.

Abundance ratios would be a better tool in DLAs chemical evolution studies.
Unlike absolute values, abundance ratios do not strongly depend
on particular model parameters, but mainly on the stellar 
nucleosynthesis and the adopted IMF. The [$\alpha$/Fe] and [N/$\alpha$] 
ratios can be used to settle constraints on time scales for star formation 
and metal production. 
While the $\alpha$ elements are mainly produced in type II 
Supernovae (SNe II) in short time scales, the Fe-peak elements are produced 
in type Ia Supernovae (SNe Ia) in a longer time scale. 
A slow (long time scale) star formation 
correspond to nearly solar values of [$\alpha$ /Fe] while 
suprasolar values may indicate shorter time scales for the star formation,
which is dominated by SNe II. The [N/$\alpha$] ratio can also indicate
the star formation time scale, because the ejection of nitrogen 
(mainly produced in intermediate mass stars)
into the interstellar medium (ISM) is delayed 
with respect to that of the $\alpha$ elements. 
There is a controversy, though, 
about the observed pattern of the [$\alpha$/Fe] ratio in DLAs. 
Some authors (Pettini et al. 2000a, Ellison et al. 2001a)
detect an [$\alpha$/Fe] enhancement
similar to that of the metal-poor stars of our Galaxy,
although the enhancement levels are typically lower in DLAs.
In contrast, other works find nearly solar [$\alpha$/Fe] ratios,
as expected from the chemical evolution of dwarf galaxies
(Centuri\'on et al. 2000, Molaro et al. 2001).

The discrepancy between the values of the [$\alpha$/Fe] ratio
determined by different authors
may be related to the way they consider dust depletion.
It has long been realized that dust depletion
is a major problem concerning the \chev\ of DLAs,
because the observed abundances probe only the gas phase of the DLA ISM.
Pettini et al. (1994) demonstrated that the 
majority of DLAs exhibit an enhancement of Zn relative to Cr,
which they interpreted as evidence of dust depletion.
In addition,
DLA surveys generally use QSOs selected from magnitude limited samples,
in which QSOs that lie behind dusty, and presumably metal-rich DLAs would
be under-represented (Pei \& Fall 1995).
One must know the depletion  
fraction of each element in each system in order to determine its real  
abundance pattern; the observed abundance refers only to the fraction of the element in gaseous form. 
Several authors have proposed different approaches to this problem,
but there is not yet consensus about the procedure to correct for
dust depletion.
Pettini et al. (2000a) applied an uniform dust depletion correction to
refractory elements in systems for which dust/gas ratio is less than 50 \%.
Savaglio et al. (1999) compared the DLAs depletion patterns to those  
of our own Galaxy in order to elaborate a correction for each system
separately. Vladilo (1998, 2002b), on the other hand, suggested 
more elaborated methods for dust correction based on the dust 
chemical composition and intrinsic abundances of two elements (Zn and Fe)
with different dust depletion properties.
Previous works (Vladilo 1998) assumed an intrinsic solar Zn/Fe ratio,
but later works (Vladilo 2002b) considered 
a possible nucleosynthetic enhancement of this ratio.
Recovering the intrinsic abundaces of the ISM is a very important goal,
but applying dust corrections is a complex task, and, therefore,
to take into account the uncertainties in the dust correction process,
it is better to consider a variety of dust correction models,
guided by some sensible assumptions about the chemistry of
dust production in the ISM, and considering a range of 
nucleosynthetic contributions to the Zn/Fe ratio.

Constraining the nature of the DLAs is crucial for
understanding the passage of a gas rich universe to 
the present day universe, where
the directly observable baryonic content of galaxies is concentrated in stars.
The  DLAs are the dominant reservoir of neutral baryons at
every observable epoch and are identified as
the progenitors of modern galaxies,
due to the correspondence of the
inferred baryonic mass density of DLAs at $z \sim 2$ and the
mass density in stars today (Storrie-Lombardi \& Wolfe 2000).
Since DLAs are the most directly observable baryonic mass systems at high 
redshift, their properties are an important constraint for theories of \gfor.
One of the most exciting aspects of the DLAs is the fact that
they provide a means for detecting
galactic systems in the early stages of evolution.
If DLAs are associated to one particular class of objects
--- e.g. disks, spheroids, galactic halos --- their study
allow us to follow the \ev\ of this population.

Nevertheless, the nature of the DLAs remains an intriguing problem.
Results from early DLA surveys (Wolfe et al. 1986, Lanzetta et al. 1991)
suggested that these systems could be the progenitors of the present 
day spirals. This hypothesis is supported by the kinematics of metal 
absorption lines exhibiting leading edge asymmetry,
which could be explained by large rotating disks
(Prochaska \& Wolfe 1997, 1998).
However, Ledoux et al. (1998) found that the correlation of
the broadening of the absorption lines
$\Delta V$ with the asymmetry of the lines,
expected in the disk interpretation,
holds only for $\Delta V < 150$ \kms,
while the kinematics of the high $\Delta V$ DLAs
is consistent with random motions,
suggesting the merging of sub-systems, as expected in CDM models.
In fact, the predictions of CDM cosmologies
are often incompatible with the large disk interpretation.
In hierarchical CDM models (Kauffmann 1996), high 
redshift galaxies should be relatively compact and underluminous in stark 
contrast with the presence of massive disks at $z\sim 3$.
The semi-analytical models of Mo, Mao \& White (1998),
in which the DLAs are identified to
centrifugally supported disks at the centre of dark matter halos,
predict that most DLAs should be low surface brightness disks,
with a median $V_{rot}<100$ \kms,
pointing to low surface brightness galaxies (LSBs)
rather than large disks as the main candidates for DLAs.
On the other hand, the DLA data can be used to constrain
semi-analytical models of galaxy formation,
which include a variety of physical processes
(Kauffmann \& Haelnelt 2000, Sommerville, Primack \& Faber 2000).
In addition, the CDM models have shown that the
rotating disk interpretation for DLAs is not unique and that
the kinematical data can also be reproduced by
infalling sub-galactic clumps in collapsing dark matter halos
(Haehnelt, Steinmetz \& Rauch (1998).
All these studies are consistent with 
a multi-population or multi-component scenario for the nature of DLAs.
A multi-component scenario is suggested by
the kinematics of the lines of high ions (CIV and SiIV),
which indicate that the high ion subsystem is kinematically distinct from
the low ion subsystem, although both subsystems are interrelated,
as would be the case of a disk at the centre of a dark matter halo,
filled with ionized gas radially infalling toward the disk
(Wolfe \& Prochaska 2000a,b).

The multi-population scenario, in which DLAs
trace different types of galaxies, and not a single galaxy class,
receives support from imaging studies of low redshift ($z_{abs} \la 1$) DLAs
(Le Brun et al. 1997; Pettini et al. 2000a; Turnshek et al. 2001),
which reveal a variety of morphologies for DLA host candidates:
spirals, compact dwarfs, low surface brightness galaxies.
The luminosities of these objects also span a wide range:
from a $\sim 0.01 L_B^*$ LSB (Bowen et al. 2001) 
to a $3 L_B^*$ spiral (Le Brun et al. 1997).
At higher redshift, the data on DLA hosts are scarce.
The three hosts of DLA at $\ga 2$ studied by M\o ller et al. (2002) 
are typically sub-$L^*$ objects,
and one of them has an irregular morphology.

The disagreement between some chemical evolution studies of DLAs
is also suggestive of a multi-population scenario for DLAs.
The controversy about the enhancement of the $\alpha$ elements in DLAs
illustrates this point.
Some studies (Pettini et al. 1997) found that
the observed abundance pattern resembles 
those of the Galactic thin and thick disk,
giving support to the disk scenario for DLAs.
On the other hand, other authors (Molaro et al. 2001, Centuri\'on et al. 2000)
claimed that chemical enrichment of DLAs is characterized
by an episodic or quiescent star formation regime
similar to predictions of dwarf galaxy models 
(Matteucci et al. 1997).

In order to understand the nature of the DLAs,
there have been several works comparing observations of DLAs
with the predictions of chemical evolution models.
Matteucci et al. (1997) studied the $\alpha$/Fe and N/O ratios observed in DLAs
by means of chemical evolution models for starburst galaxies 
and the solar neighbourhood,
and concluded that the most promising models to explain 
the observed abundances are those 
successfully describing dwarf irregular galaxies,
which assume short but intense bursts of \sfor.
Jimenez et al. (1999) coupled a disk \for\ model to a \chev\ model
to investigate low surface brightness galaxies and
high surface brightness galaxies, and conclude that LSBs could
dominate the high z DLA population.
Prantzos \& Boissier (2000) and Hou, Boissier \& Prantzos (2001)
used chemical evolution models for disk evolution and showed how 
observational constraints on the metal column density of DLAs lead naturally to
the observed mild redshift evolution of the abundances of several elements.
Cen et al. (2002) used cosmological hydrodynamic simulations
to compute the metallicity evolution of DLAs.
In their models, the observed slow \ev\ of the DLA metallicity
is explained by the steady conversion of the
highest metallicities sites to galaxies, thus depleting this class,
while all the lower metallicity systems show, individually, an
increase in metallicity.
Calura, Matteucci \& Vladilo (2002) compared the observed chemical abundances
to predictions of one-zone or multi-zone models of \chev,
and found that DLAs can be  explained
either by disks of spirals observed at large galactocentric distances,
or by LMC-like irregular galaxies,
or by starburst dwarf irregulars observed  at different times after the last 
burst of star formation.
Finally, we should mention the SPH simulations of
Nagamine, Springel \& Hernquist (2003),
based on a conservative entropy formulation,
and which include feedback by supernovae and galactic winds.
The results of these simulations agree well with observations at $z=3$
of the DLA abundance, their total neutral mass density, 
and column density distribution.

In the present work we compare the observations of DLAs with the predictions of
three classes of models: i) one-zone models for dwarf galaxies; ii) multi-zone
models for disk galaxies; iii) chemodynamical models for spheroids. 
The comparison with the results of the chemodynamical model
represents a novelty of the present work, since, differently from the
pure chemical evolution models, the \chdyn\ models allow to calculate
the gas flow in a consistent way instead of being introduced ad hoc.
The connection between gas flows and star formation
provides a natural explanation for the variety of chemical properties of DLAs.
In addition, in the dwarf galaxy scenario for DLAs, the bursts
occurs naturally, and one does not need additional model parameters
characterizing the burst.
Finally, the chemodynamical model,
within the scenario in which the DLAs are dwarf galaxies,
provides the properties of the DLA as a function of the 
\gal\ impact parameter, 
which is not possible in the one-zone model.
In this sense, the present work
complements previous works, which, within the disk scenario for disks,
study the properties of DLAs as a function of the galactocentric distance
(Prantzos \& Boissier 2001, Hou et al. 2001, Calura et al. 2003).

In section 2, we perform a statistical robust analysis
of the available chemical abundance data for DLAs,
and correct the abundances for dust depletion.
Section 3 presents the classes of chemical evolution models used here.
In section 4, we apply  models of distinct classes
to examine several scenarios for DLAs.
In Section 5, we summarise our results and,
within the framework of the chemodynamical model of DLAs,
we discuss the connection between gas flows and chemical evolution of galaxies,
the epoch of galaxy formation, 
the \ev\ of the neutral gas content in DLAs,
the metal content of the high-redshift Universe,
and the relation between DLAs and LBGs.
In this paper we adopt
an $\Omega_m = 0.3$, $\Omega_{\Lambda}= 0.7$, $H_0=70$ \kms\ Mpc$^{-1}$
cosmology (the corresponding age of the universe is 13.47 Gyr).

\begin{table*}
\caption{\small Chemical Abundances in Damped Lyman $\alpha$ Systems.}
\begin{tiny}
\begin{tabular}{lllccccccccc}
\hline\hline\noalign{\smallskip}
QSO &$z_{abs}$ &log N(HI) &[Zn/H] &[Cr/H] &[Fe/H] &[Si/H] &[S/H]
&[O/H] &[N/H] &Ref.$^a$\\
\noalign{\smallskip}
\hline
Q0000-2620 &3.390 &21.41$\pm 0.08$ &-2.07$\pm 0.10$ &-2.00$\pm 0.09$
&-2.04$\pm 0.09$ &-1.91$\pm 0.08$ &-1.91$\pm 0.09$  &-1.73$\pm 0.15$
&-2.61$\pm 0.12$ &6\\  
QXO0001 &3.000 &20.70$\pm$0.05 &-- &-- &$<$-1.11 &-1.81$\pm 0.05$ 
&-- &-1.67$\pm 0.06$ &-3.22$\pm 0.06$ &10\\
Q0013-004 &1.973 &20.83$\pm$0.04 &-0.75$\pm 0.06$ &-1.52$\pm 0.04$
&-1.51$\pm 0.05$ &-0.95$\pm 0.05$ &-0.74$\pm 0.04$ &-- &-- &18\\  
BR0019-15 &3.439 &20.92$\pm$0.10 &-- &-- &$>$-1.63
&-1.06$\pm 0.10$ &-- &-- &-- &8\\  
Q0100+13 (PHL957) &2.309 &21.40$\pm$0.05 &-1.62$\pm 0.06$ &-1.69$\pm 0.05$
&-1.93$\pm 0.05$ &-1.37$\pm 0.05$ &-1.40$\pm 0.05$ &$>$-2.70
&-1.96$\pm 0.35$ &8,4\\
Q0149+33 &2.141 &20.50$\pm$0.10 &-1.67$\pm 0.10$ &-1.45$\pm 0.10$
&-1.77$\pm 0.10$ &-1.49$\pm 0.10$ &-- &$<$0.56 &-- &8\\
Q0201+1120 &3.386 &21.26$\pm$0.08 &-- &-- &-1.40$\pm 0.21$ &--
&-1.25$\pm 0.18$ &-- &-1.86$\pm 0.27$ &5\\
Q0201+36 &2.463 &20.38$\pm$0.04 &-0.29$\pm 0.10$ &-0.80$\pm 0.05$
&-0.87$\pm 0.05$ &-0.41$\pm 0.05$ &-- &-- &$>$-1.31 &8,10\\
Q0216+0803 &1.769 &20.00$\pm$0.18 &$<$-0.33 &$<$-0.63 &-0.97$\pm 0.20$
&-0.67$\pm 0.21$ &-- & -- &-- &2\\  
Q0216+0803 &2.293 &20.45$\pm$0.16 &$<$-0.31
&$<$-0.75 &-1.06$\pm 0.18$ &-0.56$\pm 0.16$  &-- &-- &-- &2\\
J0255+00 &3.253 &20.70$\pm$0.10 &-- &-- &-1.44$\pm 0.10$ &-0.94$\pm 0.11$
&-- &-- &-- &8\\
J0255+00 &3.915 &21.30$\pm$0.05 &-- &-- &-2.05$\pm 0.10$ &$\geq$-2.67
&-1.78$\pm 0.05$ &$\geq$-2.87 &-- &8\\
Q0302-223 &1.009 &20.36$\pm$0.11 &-0.58$\pm 0.12$ &-0.97$\pm 0.12$
&-1.19$\pm 0.12$ &-0.74$\pm 0.12$ &-- &-- &-- &1\\
BRJ0307-4945 &4.466 &20.67$\pm$0.09 &-- &-- &-1.97$\pm 0.19$ &-1.54$\pm 0.11$
&-- &-1.50$\pm 0.19$  &-3.02$\pm 0.15$ &7\\  
Q0336-01 &3.062 &21.20$\pm$0.10 &-- &-- &-1.79$\pm 0.10$ &$\geq$-1.62
&-1.41$\pm 0.10$ &$\geq$-1.00 &$>$-2.09 &8,10\\
Q0347-38 &3.025 &20.62$\pm$0.01 &-1.50$\pm 0.05$ &-1.88$\pm 0.04$
&-1.69$\pm 0.01$ &-1.17$\pm 0.03$  &$<$-1.07 &-0.73$\pm 0.06$ &-1.67$\pm 0.06$
&3,10\\
Q0454+039 &0.860 &20.69$\pm$0.06 &-1.03$\pm 0.12$ &-0.89$\pm 0.08$
&-1.02$\pm 0.08$ &-0.80$\pm 0.12$ &-- &-- &-- &1\\
PKS0458-020 &2.040 &21.65$\pm$0.09 &-1.19$\pm 0.09$ &-1.50$\pm 0.10$
&-1.77$\pm$0.10 &$>$-1.17 &-- &-- &-- &8\\
PKS0528-2505 &2.141 &20.70$\pm$0.08 &$<$-1.09 &-1.29$\pm$0.09
&-1.26$\pm$0.27
&-1.00$\pm$0.09 &-1.07$\pm$0.09 &-- &-2.05$\pm$0.11 &2,19 \\  
PKS0528-2505 &2.811 &21.20$\pm$0.10 &-0.78$\pm$0.12 &-1.24$\pm$0.16
&-1.25$\pm$0.15
&-0.76$\pm$0.11 &-0.84$\pm$0.10 &$>$-2.00 &$\leq$-1.47 &2,19\\
Q0551-366 &1.962 &20.50$\pm$0.08 &-0.15$\pm0.09$ &-0.92$\pm$0.10
&-0.95$\pm0.09$ &-0.44$\pm0.10$  &-0.32$\pm0.14$ &-- &-- &14\\
HS0741+4741 &3.017 &20.48$\pm$0.10 &-- &-- &-1.93$\pm$0.10
&-1.69$\pm$0.10 &-1.68$\pm$0.10 &$\geq$-1.46 &-2.44$\pm$0.10 &8,10\\
Q0836+11 &2.465 &20.58$\pm$0.08 &$\leq$-1.13 &$\leq$-1.35 &-1.40$\pm 0.10$
&-1.15$\pm 0.11$ &$\leq$-1.12 &$\geq$-1.82 &-- &8\\
Q0841+129 &2.375 &20.95$\pm$0.09 &-1.51$\pm$0.10 &-1.55$\pm$0.10
&-1.58$\pm$0.10 &-1.27$\pm$0.10 &-1.38$\pm$0.09 &$>$-1.97
 &-2.26$\pm$0.10 &8,19\\
Q0841+129 &2.476 &20.78$\pm$0.10 &-1.40$\pm$0.10 &-1.61$\pm$0.10
&-1.75$\pm$0.10 &1.39$\pm$0.14 &1.39$\pm$0.10 &$>$-1.92 
 &-2.59$\pm$0.10 &8,19\\
Q0930+2858 &3.235 &20.18$\pm$0.10 &-- &-- &-- &$\geq$-1.80
&-1.71$\pm$0.15 &$\geq$-2.38 &-2.29$\pm$0.15 &4\\
BRI0951-0450 &3.857 &20.60$\pm$0.10 &-- &-- &-2.00$\pm$0.10 &$>$-1.53
&-- &-- &-- &8\\
BRI0951-0450 &4.203 &20.40$\pm$0.10 &-- &-- &$<$-2.59 &-2.62$\pm$0.10
&-- &-- &-- &8\\
PSS0957+33 &4.178 &20.50$\pm$0.10 &-- &-- &-1.87$\pm 0.11$ &-1.50$\pm 0.10$
&-1.31$\pm 0.12$ &$\geq$-1.90 &-- &8\\
PSS0957+33 &3.279 &20.32$\pm$0.08 &$\leq$-0.86 &-- &-1.45$\pm 0.08$
&-1.00$\pm 0.10$ &-- &-- &-- &8\\
Q1055+4611 &3.317 &20.34$\pm$0.10 &-- &-- &-- &$\geq$-1.61
&$\leq$-1.19 &$\geq$-1.97 &$\leq$-2.18 &4\\
HE1104-1805 &1.662 &20.85$\pm$0.01 &-1.04$\pm$0.01 &-1.47$\pm$0.01
&-1.58$\pm0.02$
&-1.03$\pm 0.02$ &-- &-0.79$\pm$0.20 &-1.74$\pm$0.16 &15\\
BRI1108-07 &3.608 &20.50$\pm$0.10 &-- &-- &-2.12$\pm 0.10$ &-1.80$\pm 0.10$
&-- &$\geq$-2.37 &-- &8\\
BR1117-1329 &3.350 &20.84$\pm$0.13 &-1.18$\pm$0.13 &-1.36$\pm$0.13
&-1.51$\pm$0.13 &-1.26$\pm$0.13 &-- &$\geq$-1.25 &$\leq$-2.24 &13\\
BRI1202-0725 &4.383 &20.60$\pm$0.13 &-- &-- &-2.22$\pm$0.16
&-1.77$\pm$0.12 &-- &$>$-1.84 &$\leq$-2.23 &2,4\\
Q1210+17  &1.892 &20.60$\pm$0.10 &-0.90$\pm$0.10 &-1.00$\pm$0.10
&-1.15$\pm$0.12 &-0.87$\pm$0.10
&-- &-- &-- &8\\
GC1215+3322 &1.999 &20.95$\pm$0.07 &-1.29$\pm$0.08 &-1.52$\pm$0.10
&-1.70$\pm$0.08 &-1.48$\pm$0.07 &-- &$>$-2.56 &-- &8\\
Q1223+17 &2.466 &21.50$\pm$0.10 &-1.62$\pm$0.10 &-1.68$\pm$0.10
&-1.84$\pm$0.10
&-1.59$\pm$0.10 &-- &$\geq$-2.76 &-2.60$\pm$0.18 &8,10\\
Q1232+0815 &2.338 &20.80$\pm$0.10 &-- &-- &-1.59$\pm$0.13
&-1.18$\pm$0.13 &-1.17$\pm$0.14 &-- &-2.10$\pm$0.13 &19\\
MC1331+170 &1.776 &21.18$\pm$0.04 &-1.30$\pm$0.05 &-1.97$\pm$0.04
&-2.06$\pm$0.05 &-1.45$\pm 0.10$ &-- &-- &-1.82$\pm$0.10 &8,10\\
BRI1346-0322 &3.736 &20.72$\pm$0.10 &-- &-- &$\leq$-2.09 &-2.33$\pm 0.10$
&--
&$\geq$-2.44 &-- &8\\
Q1351+318 &1.149 &20.23$\pm$0.10 &-0.38$\pm 0.17$ &-0.94$\pm 0.17$
&-0.99$\pm 0.12$  &-0.56$\pm 0.17$ &-- &-- &-- &16\\  
Q1354+258 &1.420 &21.54$\pm$0.06 &-1.63$\pm 0.16$
&-1.82$\pm 0.09$ &-2.02$\pm 0.08$  &-1.74$\pm 0.15$
&-- &-- &-- &16\\
Q1409+095 &2.456 &20.54$\pm$0.04 &-- &-- &-2.30$\pm 0.02$  &-2.01$\pm 0.02$
&-- &-2.13$\pm 0.06$ &$\leq$-3.28 &12\\  
Q1425+6039 &2.827 &20.30$\pm$0.04 &-- &-- &-1.33$\pm 0.04$
&$>$-1.03 &-- &-- &-1.53$\pm 0.04$ &10\\  
GB1759+7539 &2.625 &20.80$\pm$0.10 &$>$-1.78 &-1.26$\pm$0.10 &-1.21$\pm$0.10
&-0.82$\pm 0.10$ &-0.76$\pm 0.10$ &$>$-1.28 &-1.53$\pm 0.02$ &8,10\\  
Q1946+7658 &2.843 &20.27$\pm$0.06 &-- &-- 
&-2.53$\pm 0.06$ &-2.23$\pm 0.06$ &$<$-1.98 &-2.19$\pm 0.06$  
&-3.61$\pm 0.07$ &8,10\\
Q2206-19 &1.920 &20.65$\pm$0.07 &-0.41$\pm 0.07$ &-0.68$\pm 0.07$
&-0.86$\pm 0.07$ &-0.42$\pm 0.07$ &-- &-- &-- &8\\
Q2206-199N &2.076 &20.43$\pm$0.06 &$<$-1.90 &$<$-2.19  &-2.61$\pm 0.06$
&-2.31$\pm 0.07$  &-- &-1.94$\pm0.09$ &$\leq$-3.49 &8,12\\
Q2212-1626 &3.662 &20.20$\pm$0.08 &-- &-- &$\leq$-1.77
&-1.91$\pm 0.08$ &-- &$\geq$-2.18 &$\leq$-2.55 &2,4\\  
LBQS2230+0322 &1.864 &20.85$\pm$0.08 &-0.72$\pm 0.10$
&-1.12$\pm 0.10$ &-1.17$\pm 0.09$  &-0.75$\pm 0.10$
&-- &-- &-- &8\\
Q2231-0015 &2.066 &20.56$\pm$0.10 &-0.88$\pm 0.10$ &-1.06$\pm 0.10$
&-1.40$\pm 0.10$ &-0.87$\pm 0.10$ &-- &-- &-- &8\\
Q2237-0608 &4.080 &20.52$\pm$0.11 &-- &-- &-2.17$\pm 0.16$
&-1.81$\pm 0.11$ &-- &--  &$\leq$-2.17 &2,4\\
HE2243-6031 &2.330 &20.67$\pm$0.02 &-1.12$\pm 0.05$ &-1.12$\pm 0.09$
&-1.25$\pm 0.02$ &-0.87$\pm 0.04$ &-0.85$\pm 0.03$ &-0.68$\pm 0.21$
&-1.72$\pm 0.02$ &11\\
B2314-409 &1.874 &20.10$\pm$0.20 &$<$-1.21 &$<$-1.62 &-1.87$\pm 0.24$
&-1.87$\pm 0.22$ &$<$-1.66 &$\geq$-2.05$\pm$0.22 &-- &9\\
B2314-409 &1.857 &20.90$\pm$0.10 &-1.04$\pm 0.14$ &-1.21$\pm 0.13$
&-1.32$\pm 0.14$ &-1.05$\pm 0.14$ &-1.00$\pm 0.18$ &-- &-- &9\\
Q2343+1230 &2.431 &20.35$\pm$0.05 &-0.57$\pm 0.06$
&-- &-1.19$\pm 0.06$ &-0.65$\pm 0.06$
&-0.83$\pm 0.08$ &$\geq$-1.73 &-1.60$\pm 0.10$ &4,17\\
Q2344+124 &2.538 &20.36$\pm$0.10 &-- &-- &-1.83$\pm 0.10$ &-1.74$\pm 0.10$
&$<$-1.36 &$>$-2.07 &-2.51$\pm 0.10$ &8,10\\  
Q2348-01 &2.615 &21.30$\pm$0.10 &$\leq$-2.10 &-2.30$\pm 0.12$
&-2.23$\pm 0.13$ &-1.97$\pm 0.12$ &-- &-- &-- &8\\
Q2348-01 &2.426 &20.50$\pm$0.10 &-- &$\leq$-1.46 &-1.39$\pm 0.10$
&-0.69$\pm 0.10$ &-- &-- &-- &8\\
Q2348-14 &2.279 &20.56$\pm$0.07 &-- &-- &-2.24$\pm 0.10$
&-1.92$\pm 0.10$ &-2.03$\pm 0.10$ &-- &$<$-3.26 &8,10\\
Q2359-02 &2.095 &20.70$\pm$0.10 &-0.77$\pm 0.10$ &-1.55$\pm 0.10$
&-1.65$\pm 0.10$ &-0.78$\pm 0.10$ &-- &-- &-- &8\\
Q2359-02 &2.154 &20.30$\pm$0.10 &$<$-1.07 &-1.37$\pm 0.10$
&-1.88$\pm 0.10$ &-1.58$\pm 0.10$ &-- &-- &-- &8\\  
\hline
\hline  
\end{tabular}
\end{tiny}
\begin{small}
$^a$References: 1) Pettini et al. 2000a; 2) Lu et al.  1996; 
3) Levshakov et al. 2002;
4) Lu et al. 1998b; 5) Ellison et al. 2001a; 6) Molaro et al. 2001;
7) Dessauges-Zavadsky et al. 2001; 
8) Prochaska et al. 2001; 9) Ellison $\&$ Lopez 2001, 2002;
10) Prochaska et al. 2002b; 11) Lopez et al. 2002; 12) Pettini et al. 2002;
13) P\'eroux et al. 2002b; 14) Ledoux et al. 2002; 15) Lopez et al. 1999,
16) Pettini et al. 1999;
17) Dessauges-Zavadsky et al. 2002; 18) Petitjean et al. 2002;
19) Centuri\'on et al. 2003.
\end{small}
\end{table*}

\section{Analysis of chemical abundance data of DLAs}

In the past few years, the intense effort devoted to 
the spectroscopy of DLAs resulted in
a large amount of chemical abundances derived for these systems.
A compilation of chemical abundance measurements in DLAs 
based on the literature through early 2003 is shown in Table 1.
All of the measurements in the table are from high (echelle) resolution
obtained with 8-10 m telescopes.
We consider abundance measurements of the following elements:
N; Cr, Fe and Zn for the Fe-peak elements; O, Si and S for $\alpha$-elements.
Mg is not used as an $\alpha$-element due to the difficulty in its 
determination and the uncertainties in its chemical evolution.
Table 1 exhibits
the HI density column N(HI) (\cc) in logarithm form,
and the chemical abundances in the conventional notation 
[X/H]=log(X/H)$_{obs}$-log(X/H)$_{\odot}$. 
In this paper, the reference solar abundances (X/H)$_{\odot}$ are
the N and O photospheric values of Holweger (2001), and
the meteoritic values of Grevesse \& Sauval (1998) for the other elements. 
Whenever the original paper listed in Table 1 uses a solar abundance set
different from that above,
the chemical abundances have been recomputed
according to the solar abundances adopted here.
We restrict the HI column densities to log N(HI)$\geq 20.0$, 
in order for ionization corrections to be unimportant.
Our DLA lower log N(HI) cut-off is somewhat 
below the classical (Wolfe et al. 1986) limit of log N(HI)$=20.3$.
The present work is concerned with the [$\alpha$/N] and [N/$\alpha$] ratios
as diagnostic tools of chemical evolution of the DLAs.
Therefore, Table 1 includes only DLAs with at least one 
chemical abundance measurement
(actual determination or limit) of an $\alpha$-element.
Silicon is the most readily observed $\alpha$-element,
and the Si II 1808 \AA\ transition 
is the most widely line used in the determination of the Si abundance in DLAs,
since it is usually unsaturated 
allowing one to derive reliable Si II column densities.
We then set $z=0.85$ as the low redshift cut-off of our sample,
because, at this redshift, the SiII$\lambda$1808 line
falls longward (at 3345 \AA) of the atmospheric cut-off near 3200 \AA.

\subsection{Statistical Analysis} 

The data set of Table 1 is large enough to allow for 
the use of statistical methods
in analysing abundances and abundance ratios in DLAs.
Following the approach described by Ryan et al. (1996) we make use
of robust statistical tools in order to define the best fit to the data,
employing the techniques described by Cleveland $\&$ Kleiner (1975).
Measurements for which only upper or lower limits are available are not included
in the analysis. 
We obtain three resume lines that
summarise the trend of the data and are defined as follows: MM
(midmean) - the average of all the points between and including the
quartiles of the distribution over a given range, LSMM
(lower semi-midmean) - the midmean for all the points below
the median and USMM (upper semi-midmean) - the midmean for all
the points above the median. 
If the data are scattered about the midmean according to a normal distribution,
the semi-midmeans are estimates of the true quartiles of the distribution.
In order to obtain the resume lines,
two procedures,  LOWESS and MSTAT, are applied to the data.
LOWESS is an interactive procedure that smoothes a data distribution
using a locally weighted robust regression according
to the method described by Clevend \& Devlin (1988).
This procedure reveals the trend of the distribution, 
while MSTAT analyses the spread along the correlation.
In the LOWESS procedure, local regions around each data point are
identified according to a smoothing parameter $f$ ($0 <f <1$), which  controls
the fraction of the data included in this region.
Ideally one should try to choose a value of the smoothing parameter
that is as large as possible without distorting patterns in the data.
Suggested values are in the range $0.1-0.2$;
in this paper, we use $f=0.1$.
We refer the interested reader to the recent paper by Carretta et al. (2002),
which computes summary lines similar to those used here,
and which gives some additional references on the statistical methods
and points to a code for the reader to use.

During the chemical evolution the elements of the
Fe-peak group may follow a similar abundance pattern, 
as observed in the Milk Way metal-poor stars where chromium 
and zinc have abundance values close to iron. On the other hand,
there is an overabundance of Zn compared to Cr and Fe in DLAs: 
in our sample [Zn/H]$_{\rm median}=-1.04$ while
[Fe/H]$_{\rm median}=-1.58$ and [Cr/H]$_{\rm median}=-1.36$ (Figure 1). 
This difference is generally attributed to the presence of dust 
in these systems (Pettini et al. 1994, 1997a), which would also
affect the abundance of $\alpha$ elements, specially 
of silicon, which is highly depleted in the local ISM (Savage \& Sembach 1996).
Sulphur, on the contrary, is almost undepleted and could be used
as a dust-free tracer of $\alpha$ elements in DLAs. 

\begin{figure}
\centerline{
\psfig{figure=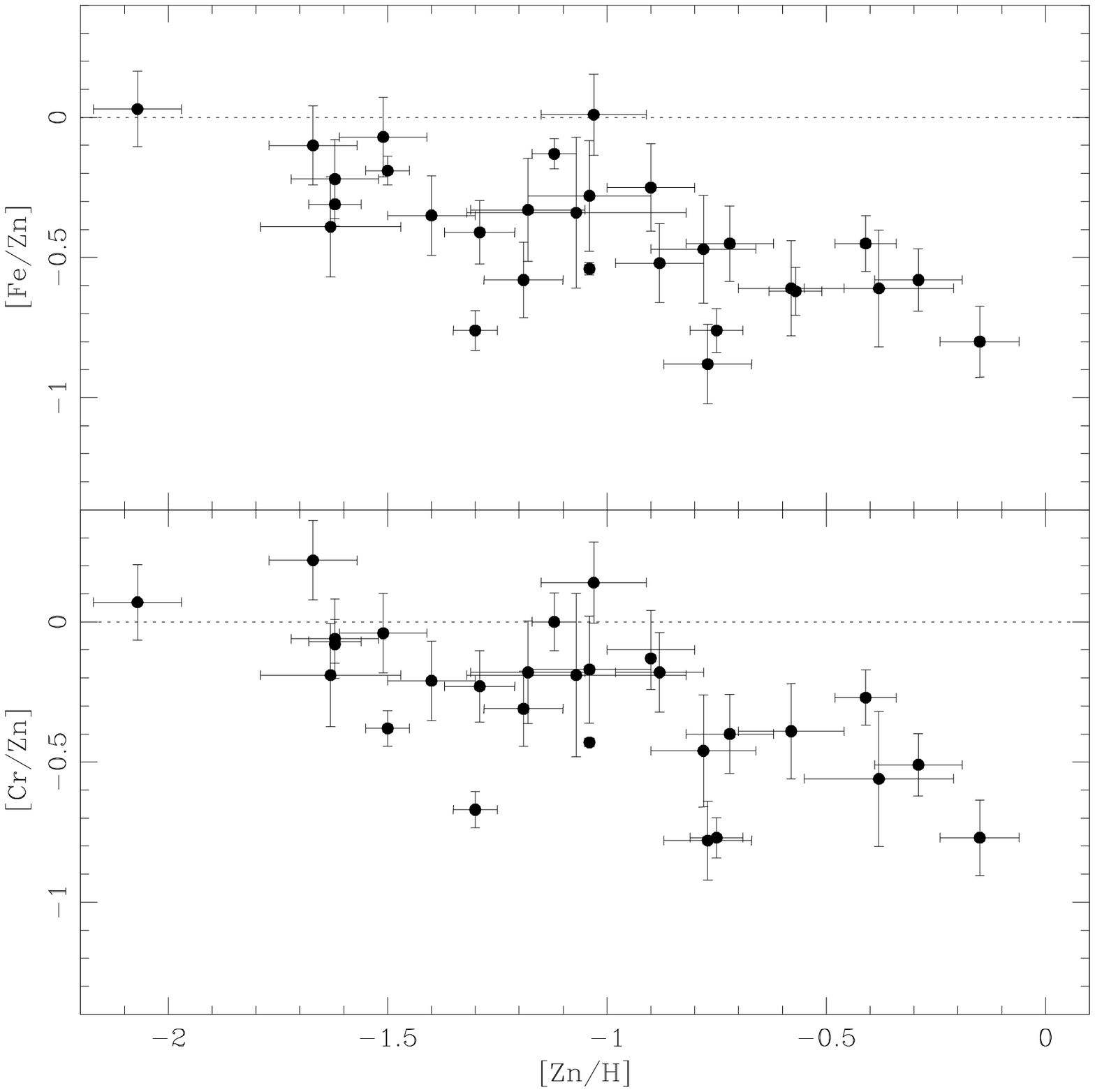,width=8.5cm,angle=0}}
\caption{[Fe/Zn] and [Cr/Zn] vs. [Zn/H] observed in DLAs.
The dotted line represents the solar value.
}
\end{figure}

\begin{figure}
\centerline{
\psfig{figure=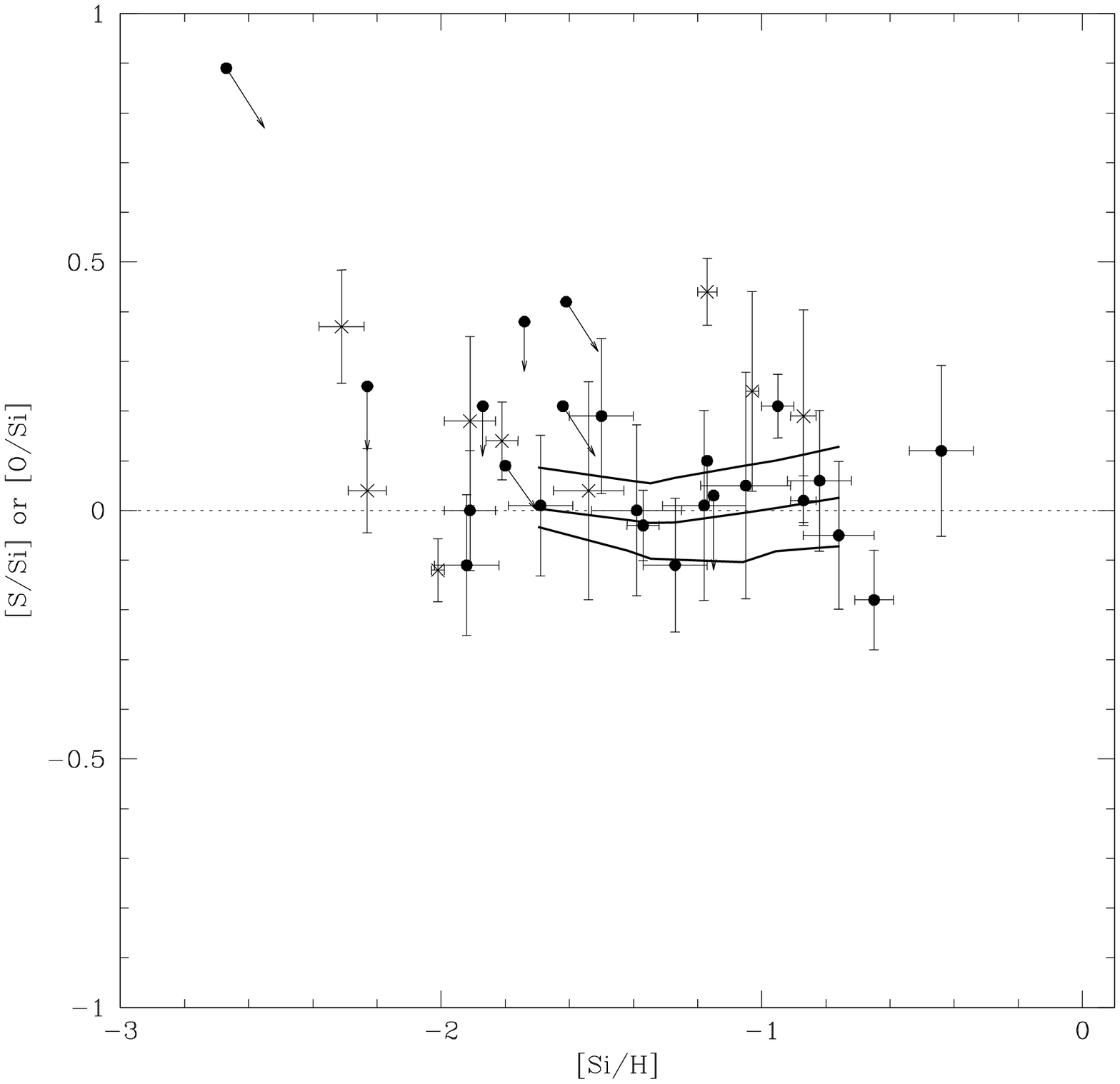,width=8.5cm,angle=0}}
\caption{[S/Si] vs. [Si/H] (filled circles) and [O/Si] vs. [Si/H] (crosses)
observed in DLAs.
The dotted line represents the solar value.
The heavy solid lines represent the robust statistical fits
to [S/Si] vs. [Si/H]:
MM (middle line), USMM (upper line) and LSMM (lower line).}
\end{figure}

However, the [S/Si] ratio in DLAs has a different behaviour
from that expected in the local ISM: the median value of the systems
that have absolute measurements of this ratio are close to solar
([S/Si]$_{\rm median}= 0.01$). 
The statistical analysis exhibits the same trend: the median
curve is on the solar value line, while the LSMM and USMM
show little dispersion around [S/Si] = 0. This fact suggests that either
the Si dust depletion is not significant in DLAs or that the S
nucleosynthesis is a result of a contribution distinct from
that in the local ISM.
If both S and Si were perfect O tracers, the [S/Si] ratio
would have intrinsic solar values and the presence of dust
(suggested in DLAs by the [Zn/Fe] and [Zn/Cr] ratios)
would imply [S/Si] $>0$.
This is not the case, as one can see from Figure 2.
DLAs exhibit both suprasolar and undersolar values of [S/Si].
Also stars of the Galactic thick disk show no sign of the $\alpha$
elements overabundance when S and Zn are used
as tracers of $\alpha$ and Fe-peak elements:
stars with $-1.2 <$ [Fe/H] $< -0.4$
have an average [S/Zn] of 0.03 ($-0.3<$ [S/Zn] $<0.3$)
(Prochaska et al. 2000).
Matteucci $\&$ Fran\c cois (1989) have already interpreted
the result of Diaz (1980) that the sulphur gradients are
flatter than the oxygen ones in spiral disks as being due to
differences between the production of O in SNe II,
which grows with the progenitor mass from
10 \msun\ to 50 \msun, and that of S, which also occurs in high
mass stars, but in a more restricted mass range though.
This interpretation is consistent with chemical evolution
models (Timmes et al. 1998) using the 
Woosley $\&$ Weaver (1995) yields for SNeII, that exhibit [S/Fe] strongly
decreasing with [Fe/H] for [Fe/H] $<$-2.
A third explanation 
(Izotov \& Thuan 1999; Izotov, Schraerer \& Charbonnel 2001)
for the variations of S/Si is
that ionization corrections for these elements are important,
since S and Si are singly ionized and could also occur in HII regions
(Viegas \& Gruenwald 1991);
in contrast, O is mostly neutral in H I gas.
All these caveats should be kept in mind when using S as
a proxy of $\alpha$ elements.
Note, however, that the [O/Si] ratio in the sample has
suprasolar values, [O/Si]$_{\rm median}= 0.18$,
which is well explained by the dust depletion scenario, 
since O is essentially undepleted in the ISM.
Unfortunately, in most cases the abundance of O is
very uncertain because it is estimated from saturated lines;
in this case, we use S as a proxy of O.

\subsection{Dust Correction}

The Si and Fe abundances are particularly important
in the study of the [$\alpha$/Fe] pattern,
since Si and Fe are the elements
tracing the $\alpha$ and iron peak group elements, respectively,
with the largest number of abundance measurements for DLAs.
However, since Fe and Si are refractory elements,
they are expected to be depleted into grains.
Ideally, one would like to gauge the level of dust depletion,
and to apply dust corrections to the observed abundances
in order to recover the intrinsic ISM (dust and gas phases) abundances.

In DLA studies, the classical indicator of dust depletion has been the 
Zn/Fe (or Zn/Cr) ratio.  Pettini et al. (1994) demonstrated that the 
majority of DLAs exhibit an enhancement of Zn relative to Cr.
This overabundance have interpreted by them as an indication of dust depletion
because Zn tends to track Fe in stars of all metallicity 
(Sneden, Gratton,  \& Crocker 1991),
whereas Zn/Fe is enhanced in dust depleted regions of the ISM
because Zn is mildly refractory while Fe is heavily depleted
(Savage \& Sembach 1996, hereafter SS96)

However, recent stellar analyses have
re-examined the Zn/Fe ratio in metal-poor and thick disk stars 
and have found enhanced Zn/Fe ratios 
(Johnson 1999, Prochaska et al. 2000, Umeda \& Nomoto 2002).
These surveys find an enhancement
of $\approx +0.1$ in the Galactic thick disk stars ([Fe/H]~$\approx -0.5$)
(Prochaska et al. 2000)
and a mean  enhancement of $\approx +0.2$~to~+0.3 
in metal-poor stars ([Fe/H]~$< -2$)
(Johnson 1999).
These results indicate the possibility of 
a nucleosynthetic contribution to the Zn/Fe enhancement of a given DLA.
On the other hand, in the Galactic ISM, 
the suprasolar [Zn/Fe] and [Si/Fe] ratios provide
evidence for dust depletion of Fe and Si ,
but the dust depletion pattern depends on the type of environment
through which the sight line passes (SS96).
SS96 consider four different dust depletion patterns 
corresponding to four types of Galactic ISM environment:
(1) cool clouds in the Galactic disk (CD), 
(2) warm clouds in the disk (WD), (3) disk plus halo clouds (WHD), 
and (4) warm halo clouds (WH).

Applying dust corrections is a very complex issue, and, therefore,
we considered a variety of dust correction models,
based on the work of Vladilo (2002a).
Following Vladilo (2002b),
we apply the dust correction models of Vladilo (2002a)
to our sample of DLAs with [Zn/Fe], [Fe/H] and [Si/H] measurements,
in order to obtain the dust-corrected  [Fe/H] and [Si/Fe] abundances.
Our prescriptions, however differ from those of  Vladilo (2002b) in that:
i) we allow for a larger scatter in the 
nucleosynthetic contributions to the Zn/Fe enhancement
by assuming a range for the intrinsic [Zn/Fe] ratio of [Zn/Fe]$_i=0.0$--0.2,
where  Vladilo (2002b) considers [Zn/Fe]$_i=0.0-0.1$;
ii) we  take as reference media the WD and the WH, 
whereas Vladilo (2002b) considers the WD and the CD.
In fact, it is not appropriate to consider the cool clouds as a reference for
the DLA environment, because the cool clouds exhibit levels of dust depletion
([Fe/Zn], [Cr/Zn], [Si/Zn]) much higher then those observed in DLAs,
and the molecular hydrogen fractions are typically low in DLAs, 
in contrast to the large amount of molecules in the cool clouds.
In addition, Savaglio, Panagia \& Stiavelli (2000) have found,
for 37 DLAs, that the majority ($\sim 60$\%) show a WH--like depletion
pattern, then $\sim 30$\% and $\sim 10$\% have a WHD--like and
WD--like depletion pattern, respectively.
Finally, Hou et al. (2001) have found evidence for a dust depletion,
for DLAs with large metal column density ([Zn/H]$+\log$N(H I)$>19.8$),
at a level ($\sim 0.4$ dex) comparable to that observed in the WH clouds. 

We summarize below the dust correction method of Vladilo (2002a). 
The dust correction method aims at obtaining the depletion 
of a given element X in the DLA:

\begin{equation}
\delta_X \;=\; [X/H]_{obs} \; -\; [X/H]_i
\end{equation}

\noindent
where $[X/H]_{obs}$ is the observed (gas) abundance
and $[X/H]_i$ the intrinsic ISM abundance of the DLA.
The depletion $\delta (X)$ is related to the fraction in dust
of the element X,
$f_X= N_{X,d}/N_{X,i} $, through:

\begin{equation}
f_X \;= 1-10^{\delta_X}  \; ,
\end{equation}

\noindent
where  $N_{X,d}$ is the column density of atoms X in  the dust and
$N_{X,i}$  the  column density in the ISM  (gas plus dust) of the DLA.

The fraction  $f_X$ is the relevant physical quantity,
which should scale with the amount of dust and 
with the abundance of X in dust. 
To derive this relation we express all abundances by number
and relative to the abundance of a reference element Y 
(here we adopt iron as Y).
Therefore, we define the relative abundance of X in the ISM as:

\begin{equation}
a_X  = { N_{X,i} / N_{Y,i}  } \; ,
\end{equation}

\noindent
which is related to [X/Y]$_i$ through $a_X=10^{[X/Y]_i}$.
We also define the {\it reference dust fraction}
(or {\it dust-to-metals ratio} as in Vladilo 2002a) in the ISM as $f_Y$.

We further assume that $f_X$ depends on $a_X$ and $f_Y$ as:

\begin{equation}
f_X \propto f_Y^{(1+\eta_X)} a_X\, ^{(\epsilon_X-1)} \; ,
\end{equation}

\noindent
in which $(1+\eta_X)$ and $(\epsilon_X-1)$ are the power law indices
of the power laws in $f_Y$ and $a_X$, respectively.
Eq. (4) allows us to obtain a general expression for $f_X$:

\begin{equation}
{f_X \over f_{X,m}} =
\left( {f_Y \over f_{Y,m}} \right)^{(1+\eta_X)}
\left( {a_X \over a_{X,m}} \right)^{(\epsilon_X-1)} \; .
\end{equation}

\noindent
If $f_{X,m}$ is known for some reference medium $m$, and, if
$\eta_X$ and $\epsilon_X$ are known or assumed, we can obtain $f_X$.
Here we use as reference environment $m$ the warm disk (WD) of the Galaxy.
Our values of $f_{X,m}$ correspond  to $\delta_{X,m}=-0.43$, -1.095, and -1.215,
for Si, Cr and Fe, respectively (Sembach \& Savage 1996),
and $\delta_{X,m}=-0.19$ for Zn (Roth \& Blades 1995).

In Vladilo (1998), $f_X$  was assumed to scale  linearly with $f_Y$.
However, this is correct when the dust composition does not vary
with $f_Y$, a situation described by $\eta_X=0$.
However, studies of the Galactic ISM indicate that the dust composition changes
with $f_Y$. In particular, for elements that are less depleted than the iron
(e.g. silicon) 
the fraction in dust, in general, decreases faster than that of the iron
for less depleted media, 
implying $\eta_X>1$. 

On the other hand, $\epsilon_X=0$ implies that the dust
composition is insensitive to variations of metallicity.
However, one expects that the dust fraction would increase
monotonically with the metallicity.
In a simple physical picture, the two-body process of adhesion of
an element onto dust grains should be proportional to the product of the
densities of that element and the dust grains.
If there are fewer atoms available, 
then the adhesion rate will be correspondingly lower.
$\epsilon_X=1$ follows from this picture,
implying that the dust fraction scales linearly with the metallicity
of the medium.
Values of $\epsilon_X > 1$
could occur if three-body processes are important in the chemistry of dust
production, but this is never the case in the low densities of the ISM.
In addition, $ \epsilon_X > 1$ not supported by
studies of the ISM of the SMC (Vladilo 2002a).
Therefore, the physically meaningful range for $\epsilon_X$
is $ 0< \epsilon_X \leq 1$.

The value of $\eta_X$ can be obtained from the comparison of 
knowledge of the dust depletion pattern in the local Galactic ISM,
considering two different environments $m$ and $n$.
We assume that the abundances are constant in the solar neighbourhood.
This implies that $a_{X,m}=a_{X,n}$ and,
as a consequence, the scaling law, eq. 5, can be written in a simplified form:

\begin{equation}
{f_{X,n} \over f_{X,m}}=
\left( {f_{Y_n} \over f_{Y,m}} \right)^{(1+\eta_X)} \; .
\end{equation}

Here, the warm disk and the warm halo
are the two media used in the determination of $\eta_X$.
For the warm halo,  Sembach \& Savage (1996) derive 
$\delta _{X,WH}=-0.26$, -0.51, and -0.64, for Si, Cr and Fe, respectively.
In their study of Zn depletion in the Galactic ISM,
Roth \& Blades (1995) included
one sightline of Sembach \& Savage in the WH category,
that toward HD167756,
and, in addition, the WDH class sightline toward HD149881.
They find for the sightlines toward HD167756 and HD149881,
$\delta_{Zn,WH}=-0.05$ and -0.11, respectively.
Accordingly, in this work we assume $\delta_{Zn,WH}=-0.10$.
However, assuming $\delta_{Zn,WH}=-0.05$ has little effect on our results
(see Table 2).
The corresponding values of $\eta_X$ are 0.685, 0.450 and 1.760,
for Si, Cr and Zn, respectively 
(=4.922 for Zn if $\delta_{Zn,WH}=-0.05$).

Once we have determined or estimated $\epsilon_X$ and $\eta_X$,
we can proceed in our attempt of dust correction of the abundances,
by considering the relation between the observed [X/H]$_{obs}$
and the intrinsic abundance [X/H]$_i$ 
(or $a_X=10^{[X/Y]_i}$ in the equations below),
which can be derived from Eq.s (1) and (5) above:

\begin{equation}
\left[ { {\rm X} \over {\rm H} } \right]_{obs} =
\left[ { {\rm X} \over {\rm H} } \right]_i 
+\log \left\{
1-\rho ^{(1+\eta_X)}
\, a_X\, ^{( \epsilon_ X - 1)} 
\,  f_{{\rm X},m} 
\right\}
\; ,
\end{equation}

\noindent
where $\rho=f_Y/f_{Y,m}$ is 
the normalized (to $f_{Y,m}$ of the reference medium $m$) dust-to-metals ratio.

In order to recover the intrinsic abundances we first need to determine $\rho$.
With this task, we apply Eq. (7) to the generic element X and to the
reference element Y to obtain the relation:
\begin{equation}
\rho 
-
{  
f_{X,m} \, a_X\, ^{\epsilon_X}
\over 
f_{Y,m} \, 10^{\left[ { {\rm X} \over {\rm Y} } \right]_{obs} }  
}
\rho^{( 1+\eta_X ) }
+
{
a_X - 10^{\left[ { {\rm X} \over {\rm Y} } \right]_{obs} } 
\over
f_{Y,m} \,  10^{\left[ { {\rm X} \over {\rm Y} } \right]_{obs} }  
}
=
0  \; ,
\end{equation}

If we have a measurement of [X/Y]$_{obs}$ and a guess on [X/Y]$_i$,
Eq. (8) can be solved to find the  root $\rho$.

Once $\rho$ is known, we can determine the intrinsic relative abundance 
[X/Y]$_i$ of a element X different from that used in Eq. (8) from the expression

\begin{equation}
a_X
-
\rho^{( 1+\eta_X)} \, f_{X,i} \,
 a_X\, ^{\epsilon_X} 
+  
\left( \rho f_{Y,m} - 1 \right)
 10^{\left[ { {\rm X} \over {\rm Y} } \right]_{obs}}
=0  \; ,
\end{equation}

\noindent
which is just a rearrangement of Eq. (8), with $a_X$ as the unknown variable
instead of $\rho$.

In applying the dust correction procedure of 
elemental abundances described above,
we follow Vladilo (2002a), and consider eight dust models,
based on our guesses on $\eta_X$ and [Zn/Fe]$_i$. 
We consider [Zn/Fe]$_i$ as an input parameter
to which we assign values 0.0 or +0.2 dex,
in order to take into account a possible nucleosynthetic Zn/Fe enhancement.
We adopt $\eta_X=0$ or $\eta_X=1$ as two limiting possibilities for $\eta_X$.
As $\eta_X$ is needed both for zinc in Eq. (8) in order to obtain $\rho$,
and for the generic element X in Eq. (9), we consider that possibility
that $\eta_{Zn}$ and $\eta_X$ are different. 
We have, therefore, eight possibilities:
S00, S10, S01, S11, E00, E10, E01, and E11, 
where the letter S or E denotes solar [Zn/Fe]$_i=0.0$ (S) 
or Zn enhanced [Zn/Fe]$_i=0.2$ (E).
The number code indicates the values of the pair ($\eta_{Zn}$,$\eta_X$)
= (0.0,0.0) (00), (1.0,0.0) (10), (0.0,1.0) (01), and (1.0,1.0) (11).
We should note that S00 and S10 ($\equiv$ S0) 
and S01 and S11 ($\equiv$ S1) are indistinguishable.
In the case of iron abundances,
S0 and S1 are indistinguishable,
and also E00 and E10, and E01 and E11.
For each dust correction model, 
the correction is applied to Si, Cr, Fe and Zn
in the systems for which the observed [Zn/Fe] is available. 

Table 2 shows, for each system, the observed abundance ratios
[Zn/Fe] and [Si/Fe] and the observed abundance of the iron [Fe/H],
the corrected abundances [Fe/H]$_{cor}$,
the corrected abundance ratios [Si/Fe]$_{cor}$,
for each dust correction model,
and the averaged corrected [Fe/H]$_{cor}$ and [Si/Fe]$_{cor}$,
$\langle$[Fe/H]$\rangle$ and $\langle$[Si/Fe]$\rangle$
(the average has been done over the eight possibilities
S00, S01, S10, S11, E00, E01, E10 and E11).
The error bars of
$\langle$[Fe/H]$\rangle$ and $\langle$[Si/Fe]$\rangle$
encompass the range of average values predicted by each dust model
and their respective $1 \sigma$ errors.
All models assume [Zn/H]$_{WH}=-0.1$ for the Galactic Warm Halo medium,
but adopting [Zn/H]$_{WH}=-0.05$ changes very little our results even
for the system with the largest [Zn/Fe]$_{obs}$, 
the $z_{abs}=2.095$ DLA toward Q2359-02 (see Table 2).

\begin{table*}
\begin{tiny}
\caption{\small Corrected abundance of Fe and Si.}
\begin{tabular}{llcccccccccc}
\hline\hline\noalign{\smallskip}
 & & & & &\multicolumn{5}{c}{Dust Model} & & \\
 & & & & &S0 &E00 &S0 &E00 &E10 & & \\
 & & & & &   &E11 &S1 &E01 &E11 & & \\
\hline
QSO &$z_{abs}$ &[Zn/Fe] &[Fe/H] &[Si/Fe]
&[Fe/H]$_{cor}$ &[Fe/H]$_{cor}$ 
&[Si/Fe]$_{cor}$ &[Si/Fe]$_{cor}$ &[Si/Fe]$_{cor}$
&$\langle$[Fe/H]$\rangle$ &$\langle$[Si/Fe]$\rangle$ \\
\noalign{\smallskip}
\hline
Q0000-2620 &3.390 &-0.03$\pm$0.09&-2.04$\pm$0.09& 0.13$\pm$0.07
 &-2.04$\pm$0.09&-2.04$\pm$0.09
 & 0.13$\pm$0.14& 0.13$\pm$0.13& 0.13$\pm$0.13
 &-2.04$^{+0.09}_{-0.09}$ & 0.13$^{+0.14}_{-0.14}$\\
 & & & & & 
 &-2.04$\pm$0.09
 & 0.13$\pm$0.14 & 0.13$\pm$0.13& 0.13$\pm$0.13 & & \\
Q0013-004  &1.973 & 0.76$\pm$0.07&-1.51$\pm$0.05& 0.56$\pm$0.06
 &-0.60$\pm$0.07&-0.89$\pm$0.07
 & 0.00$\pm$0.04& 0.12$\pm$0.07& 0.10$\pm$0.06
 &-0.73$^{+0.20}_{-0.22}$ & 0.07$^{+0.21}_{-0.17}$\\
 & & & & & 
 &-0.84$\pm$0.07
 & 0.01$\pm$0.10 & 0.19$\pm$0.09& 0.17$\pm$0.09 & & \\
Q0100+13   &2.309 & 0.31$\pm$0.07&-1.93$\pm$0.07& 0.56$\pm$0.06
 &-1.58$\pm$0.08&-1.82$\pm$0.07
 & 0.28$\pm$0.10& 0.46$\pm$0.11& 0.46$\pm$0.12
 &-1.70$^{+0.20}_{-0.19}$ & 0.39$^{+0.19}_{-0.21}$\\
 & & & & & 
 &-1.82$\pm$0.07
 & 0.34$\pm$0.10 & 0.47$\pm$0.10& 0.47$\pm$0.10 & & \\
Q0149+33   &2.141 & 0.10$\pm$0.06&-1.77$\pm$0.10& 0.28$\pm$0.06
 &-1.67$\pm$0.10&-1.77$\pm$0.10
 & 0.19$\pm$0.15& 0.28$\pm$0.12& 0.28$\pm$0.12
 &-1.72$^{+0.15}_{-0.15}$ & 0.24$^{+0.17}_{-0.19}$\\
 & & & & & 
 &-1.77$\pm$0.10
 & 0.20$\pm$0.15 & 0.28$\pm$0.12& 0.28$\pm$0.12 & & \\
Q0201+36   &2.463 & 0.58$\pm$0.10&-0.87$\pm$0.05& 0.46$\pm$0.06
 &-0.18$\pm$0.12&-0.46$\pm$0.10
 & 0.03$\pm$0.07& 0.15$\pm$0.10& 0.14$\pm$0.10
 &-0.31$^{+0.25}_{-0.25}$ & 0.11$^{+0.21}_{-0.18}$\\
 & & & & & 
 &-0.43$\pm$0.12
 & 0.05$\pm$0.12 & 0.21$\pm$0.11& 0.19$\pm$0.11 & & \\
Q0302-223  &1.009 & 0.61$\pm$0.09&-1.19$\pm$0.12& 0.45$\pm$0.09
 &-0.46$\pm$0.12&-0.74$\pm$0.12
 & 0.01$\pm$0.08& 0.13$\pm$0.11& 0.11$\pm$0.11
 &-0.60$^{+0.25}_{-0.27}$ & 0.08$^{+0.23}_{-0.21}$\\
 & & & & & 
 &-0.72$\pm$0.12
 & 0.02$\pm$0.15 & 0.18$\pm$0.13& 0.16$\pm$0.14 & & \\
Q0347-38   &3.025 & 0.19$\pm$0.05&-1.69$\pm$0.01& 0.52$\pm$0.03
 &-1.49$\pm$0.05&-1.69$\pm$0.02
 & 0.34$\pm$0.07& 0.52$\pm$0.04& 0.52$\pm$0.04
 &-1.59$^{+0.15}_{-0.12}$ & 0.44$^{+0.12}_{-0.16}$\\
 & & & & & 
 &-1.69$\pm$0.02
 & 0.38$\pm$0.06 & 0.52$\pm$0.04& 0.52$\pm$0.04 & & \\
Q0454+039  &0.860 &-0.01$\pm$0.13&-1.02$\pm$0.08& 0.22$\pm$0.13
 &-1.02$\pm$0.09&-1.02$\pm$0.08
 & 0.22$\pm$0.21& 0.22$\pm$0.18& 0.22$\pm$0.18
 &-1.02$^{+0.09}_{-0.09}$ & 0.22$^{+0.21}_{-0.21}$\\
 & & & & & 
 &-1.02$\pm$0.08
 & 0.22$\pm$0.21 & 0.22$\pm$0.18& 0.22$\pm$0.18 & & \\
PKS0528-2505 &2.811 & 0.47$\pm$0.14&-1.25$\pm$0.15& 0.49$\pm$0.14
 &-0.70$\pm$0.16&-0.96$\pm$0.15
 & 0.10$\pm$0.16& 0.26$\pm$0.20& 0.25$\pm$0.20
 &-0.83$^{+0.29}_{-0.29}$ & 0.20$^{+0.30}_{-0.26}$\\
 & & & & & 
 &-0.95$\pm$0.16
 & 0.16$\pm$0.21 & 0.30$\pm$0.20& 0.30$\pm$0.20 & & \\
Q0551-366  &1.962 & 0.80$\pm$0.08&-0.95$\pm$0.09& 0.51$\pm$0.09
 & 0.01$\pm$0.09&-0.29$\pm$0.09
 &-0.03$\pm$0.05& 0.07$\pm$0.08& 0.05$\pm$0.08
 &-0.13$^{+0.22}_{-0.25}$ & 0.01$^{+0.23}_{-0.23}$\\
 & & & & & 
 &-0.23$\pm$0.09
 &-0.08$\pm$0.14 & 0.12$\pm$0.13& 0.08$\pm$0.14 & & \\
Q0841+129  &2.375 & 0.07$\pm$0.08&-1.58$\pm$0.10& 0.31$\pm$0.08
 &-1.51$\pm$0.10&-1.58$\pm$0.10
 & 0.25$\pm$0.19& 0.31$\pm$0.16& 0.31$\pm$0.16
 &-1.54$^{+0.14}_{-0.14}$ & 0.28$^{+0.19}_{-0.23}$\\
 & & & & & 
 &-1.58$\pm$0.10
 & 0.25$\pm$0.19 & 0.31$\pm$0.16& 0.31$\pm$0.16 & & \\
Q0841+129  &2.476 & 0.35$\pm$0.06&-1.75$\pm$0.10& 0.36$\pm$0.06
 &-1.35$\pm$0.10&-1.60$\pm$0.10
 & 0.08$\pm$0.07& 0.23$\pm$0.09& 0.23$\pm$0.10
 &-1.47$^{+0.22}_{-0.22}$ & 0.17$^{+0.17}_{-0.16}$\\
 & & & & & 
 &-1.59$\pm$0.10
 & 0.11$\pm$0.09 & 0.25$\pm$0.09& 0.25$\pm$0.09 & & \\
HE1104-1805 &1.662 & 0.54$\pm$0.02&-1.58$\pm$0.02& 0.55$\pm$0.03
 &-0.94$\pm$0.03&-1.22$\pm$0.02
 & 0.10$\pm$0.03& 0.26$\pm$0.04& 0.24$\pm$0.04
 &-1.07$^{+0.16}_{-0.16}$ & 0.21$^{+0.15}_{-0.14}$\\
 & & & & & 
 &-1.20$\pm$0.03
 & 0.17$\pm$0.04 & 0.32$\pm$0.04& 0.31$\pm$0.04 & & \\
BR1117-1329 &3.350 & 0.33$\pm$0.07&-1.51$\pm$0.13& 0.25$\pm$0.07
 &-1.14$\pm$0.13&-1.38$\pm$0.13
 & 0.01$\pm$0.08& 0.14$\pm$0.10& 0.14$\pm$0.11
 &-1.26$^{+0.25}_{-0.25}$ & 0.08$^{+0.17}_{-0.17}$\\
 & & & & & 
 &-1.38$\pm$0.13
 & 0.02$\pm$0.10 & 0.15$\pm$0.10& 0.15$\pm$0.11 & & \\
BR1210+17   &1.892 & 0.25$\pm$0.09&-1.15$\pm$0.12& 0.28$\pm$0.09
 &-0.88$\pm$0.12&-1.10$\pm$0.12
 & 0.08$\pm$0.12& 0.23$\pm$0.13& 0.23$\pm$0.13
 &-0.99$^{+0.23}_{-0.23}$ & 0.16$^{+0.20}_{-0.20}$\\
 & & & & & 
 &-1.10$\pm$0.12
 & 0.10$\pm$0.13 & 0.24$\pm$0.13& 0.24$\pm$0.13 & & \\
GC1215+3322 &1.999 & 0.41$\pm$0.07&-1.70$\pm$0.08& 0.22$\pm$0.06
 &-1.23$\pm$0.08&-1.48$\pm$0.08
 &-0.04$\pm$0.06& 0.06$\pm$0.09& 0.06$\pm$0.09
 &-1.35$^{+0.21}_{-0.21}$ & 0.00$^{+0.16}_{-0.17}$\\
 & & & & & 
 &-1.47$\pm$0.08
 &-0.07$\pm$0.10 & 0.07$\pm$0.10& 0.07$\pm$0.10 & & \\
Q1223+17   &2.466 & 0.22$\pm$0.06&-1.84$\pm$0.10& 0.25$\pm$0.06
 &-1.60$\pm$0.10&-1.82$\pm$0.10
 & 0.08$\pm$0.08& 0.23$\pm$0.08& 0.23$\pm$0.08
 &-1.71$^{+0.21}_{-0.21}$ & 0.16$^{+0.16}_{-0.16}$\\
 & & & & & 
 &-1.82$\pm$0.10
 & 0.09$\pm$0.09 & 0.23$\pm$0.08& 0.23$\pm$0.08 & & \\
MC1331+170  &1.776 & 0.76$\pm$0.06&-2.06$\pm$0.05& 0.61$\pm$0.10
 &-1.15$\pm$0.06&-1.44$\pm$0.06
 & 0.03$\pm$0.06& 0.15$\pm$0.09& 0.13$\pm$0.09
 &-1.28$^{+0.19}_{-0.22}$ & 0.11$^{+0.26}_{-0.19}$\\
 & & & & & 
 &-1.39$\pm$0.06
 & 0.06$\pm$0.13 & 0.24$\pm$0.13& 0.22$\pm$0.13 & & \\
Q1351+318  &1.149 & 0.61$\pm$0.16&-0.99$\pm$0.12& 0.43$\pm$0.16
 &-0.26$\pm$0.18&-0.54$\pm$0.17
 & 0.00$\pm$0.13& 0.11$\pm$0.19& 0.10$\pm$0.19
 &-0.40$^{+0.32}_{-0.32}$ & 0.06$^{+0.33}_{-0.33}$\\
 & & & & & 
 &-0.52$\pm$0.18
 & 0.00$\pm$0.26 & 0.16$\pm$0.24& 0.14$\pm$0.25 & & \\
Q1354+258  &1.420 & 0.39$\pm$0.16&-2.02$\pm$0.08& 0.28$\pm$0.15
 &-1.57$\pm$0.18&-1.82$\pm$0.16
 & 0.01$\pm$0.15& 0.13$\pm$0.21& 0.12$\pm$0.22
 &-1.70$^{+0.31}_{-0.29}$ & 0.07$^{+0.31}_{-0.30}$\\
 & & & & & 
 &-1.82$\pm$0.17
 & 0.01$\pm$0.24 & 0.14$\pm$0.23& 0.14$\pm$0.24 & & \\
Q2206-19   &1.920 & 0.45$\pm$0.05&-0.86$\pm$0.07& 0.44$\pm$0.05
 &-0.34$\pm$0.07&-0.60$\pm$0.07
 & 0.08$\pm$0.06& 0.23$\pm$0.07& 0.22$\pm$0.08
 &-0.46$^{+0.20}_{-0.20}$ & 0.17$^{+0.16}_{-0.15}$\\
 & & & & & 
 &-0.59$\pm$0.07
 & 0.13$\pm$0.08 & 0.27$\pm$0.07& 0.26$\pm$0.07 & & \\
LBQ2230+0322 &1.864 & 0.45$\pm$0.09&-1.17$\pm$0.09& 0.42$\pm$0.09
 &-0.65$\pm$0.10&-0.91$\pm$0.09
 & 0.07$\pm$0.10& 0.21$\pm$0.13& 0.20$\pm$0.13
 &-0.77$^{+0.23}_{-0.22}$ & 0.16$^{+0.22}_{-0.19}$\\
 & & & & & 
 &-0.90$\pm$0.10
 & 0.11$\pm$0.14 & 0.25$\pm$0.13& 0.24$\pm$0.13 & & \\
Q2231-0015 &2.066 & 0.52$\pm$0.06&-1.40$\pm$0.10& 0.53$\pm$0.06
 &-0.79$\pm$0.10&-1.06$\pm$0.10
 & 0.10$\pm$0.07& 0.25$\pm$0.09& 0.24$\pm$0.09
 &-0.92$^{+0.23}_{-0.24}$ & 0.21$^{+0.20}_{-0.17}$\\
 & & & & & 
 &-1.04$\pm$0.10
 & 0.16$\pm$0.10 & 0.31$\pm$0.09& 0.30$\pm$0.09 & & \\
Q2243-6031 &2.330 & 0.13$\pm$0.05&-1.25$\pm$0.02& 0.38$\pm$0.04
 &-1.12$\pm$0.05&-1.25$\pm$0.02
 & 0.26$\pm$0.08& 0.38$\pm$0.04& 0.38$\pm$0.04
 &-1.18$^{+0.12}_{-0.09}$ & 0.33$^{+0.10}_{-0.14}$\\
 & & & & & 
 &-1.25$\pm$0.02
 & 0.28$\pm$0.07 & 0.38$\pm$0.04& 0.38$\pm$0.04 & & \\
B2314-409  &1.857 & 0.28$\pm$0.15&-1.32$\pm$0.14& 0.27$\pm$0.15
 &-1.01$\pm$0.16&-1.24$\pm$0.14
 & 0.06$\pm$0.18& 0.20$\pm$0.20& 0.20$\pm$0.20
 &-1.12$^{+0.28}_{-0.26}$ & 0.13$^{+0.28}_{-0.29}$\\
 & & & & & 
 &-1.24$\pm$0.15
 & 0.07$\pm$0.22 & 0.20$\pm$0.21& 0.20$\pm$0.21 & & \\
Q2343+1230 &2.431 & 0.62$\pm$0.06&-1.19$\pm$0.06& 0.54$\pm$0.06
 &-0.45$\pm$0.07&-0.73$\pm$0.06
 & 0.05$\pm$0.06& 0.19$\pm$0.08& 0.17$\pm$0.08
 &-0.58$^{+0.20}_{-0.21}$ & 0.15$^{+0.21}_{-0.15}$\\
 & & & & & 
 &-0.70$\pm$0.07
 & 0.10$\pm$0.10 & 0.26$\pm$0.09& 0.25$\pm$0.10 & & \\
Q2359-02   &2.095 & 0.88$\pm$0.06&-1.65$\pm$0.10& 0.87$\pm$0.06
 &-0.60$\pm$0.10&-0.90$\pm$0.10
 & 0.10$\pm$0.06& 0.26$\pm$0.08& 0.22$\pm$0.08
 &-0.73$^{+0.23}_{-0.26}$ & 0.24$^{+0.28}_{-0.20}$\\
 & & & & & 
 &-0.84$\pm$0.10
 & 0.21$\pm$0.10 & 0.42$\pm$0.09& 0.38$\pm$0.10 & & \\
\noalign{\smallskip} \\
Q2359-02$^a$   &2.095 & 0.88$\pm$0.06&-1.65$\pm$0.10& 0.87$\pm$0.06
 &-0.62$\pm$0.10&-0.93$\pm$0.10
 & 0.11$\pm$0.06& 0.28$\pm$0.08& 0.26$\pm$0.08
 &-0.77$^{+0.24}_{-0.26}$ & 0.26$^{+0.27}_{-0.21}$\\
 & & & & &
 &-0.89$\pm$0.10
 & 0.23$\pm$0.11 & 0.44$\pm$0.09& 0.42$\pm$0.10 & & \\
\hline\hline
\end{tabular}
\begin{small}
$^a$ The dust models assume [Zn/H]$_{WH}=-0.05$ (see text).
\end{small}
\end{tiny}
\end{table*}

\subsubsection{The [Si/Fe] ratio}

\begin{figure}
\centerline{
\psfig{figure=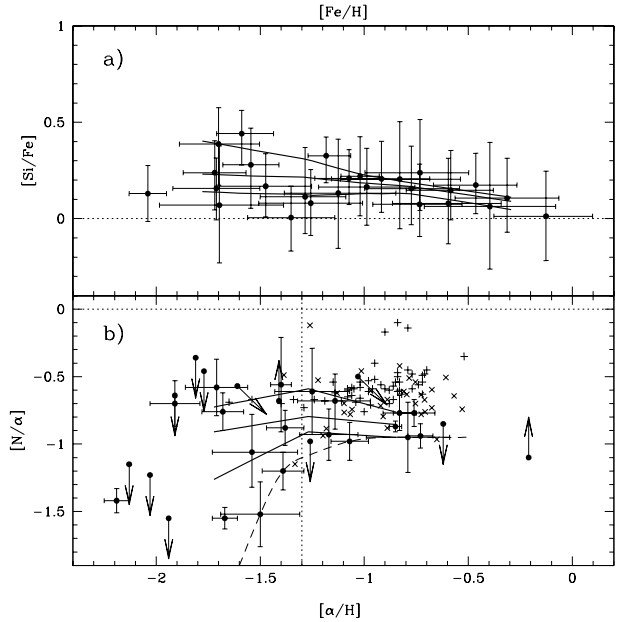,width=8.5cm,angle=0}}
\caption{
Upper panel: Dust-corrected [Si/Fe] vs. [Fe/H] for DLAs (filled circles).
Also show [S/Fe] vs. [Fe/H] data (open circles).
Lower Panel: [N/$\alpha$] vs. [$\alpha$/H] for DLAs and dwarf galaxies.
The dwarf galaxy data are [N/O] vs. [O/H]
from van Zee et al. (1997a,b) (crosses)
and Izotov $\&$ Thuan (1999) (plus signs).
For DLAs, we take as a proxy for $\alpha$ elements
O or S, or, when not available, dust-corrected Si.
The dotted horizontal lines represent the solar value,
the dotted vertical line in the bottom panel 
the value [$\alpha$/H]=-1.3 (see text),
and the heavy solid lines are the statistical fits
to [Si/Fe] vs. [Fe/H] and [N/$\alpha$] vs. [$\alpha$/H] of DLAs.
The bottom panel also shows the lower bound envelope to
the [N/$\alpha$] vs. [$\alpha$/H] distribution of DLAs (dashed line).
}
\end{figure}

The trend of the dust-corrected [Si/Fe] ratio with  metallicity ($[Fe/H]_{cor}$)
is analysed with robust statistical methods
(curves MM, USMM and LSMM in Figure 3a). 
In our sample, before any dust correction, 
the [Si/Fe] ratio is typically suprasolar, with [Si/Fe]$_{median}=0.32$.
In the corrected sample, the [Si/Fe] ratio is lowered to
[Si/Fe]$_{median}=0.16$, but remains suprasolar.
Notice that the original dust correction procedure of Vladilo (1998)
is equivalent to the case S0, which in general gives a larger
downward correction for the [Si/H] ratio.
After we perform a careful dust correction, 
we conclude that DLAs do exhibit [Si/Fe] enhancement,
and there is a trend of slight increase of [Si/H] with decreasing metallicity.
resembling that found for metal-poor stars of the Galaxy.
This suggests that a number of DLAs may have had a chemical 
evolution similar to our Galaxy,
supporting the scenario where these systems are the progenitors of disks.
However,
there are DLAs with nearly solar [Si/Fe] ratio
even at low metallicities, for which the disk picture does not apply.
This case implies a slow star formation rate,
which allows the SNe Ia to enrich the ISM with iron peak group elements. 
This behaviour is similar to that predicted by chemical 
evolution models for dwarf galaxies with bursts of star 
formation separated by quiescent periods (Matteucci et al. 1997),
which allows that class of DLAs
to be explained as (proto)dwarf galaxies.
In summary, the [Si/Fe] vs. [Fe/H] distribution of DLAs
does not exclude disks as DLA sites, 
but, if disks are present among DLA hosts, implies that they
represent one component within a multi-population scenario.

\subsubsection{The [N/$\alpha$] ratio}

The [N/$\alpha$] ratio also allows one to constrain star formation times scales
because of the delay of the injection of N with respect to
$\alpha$ elements into the ISM.
A nearly solar ratio ([N/$\alpha$] = 0) would indicate a slow star formation,
during which there is enough time for production of N.
Figure 3b show the trend [N/$\alpha$] vs. [$\alpha$/H]
for DLAs and dwarf galaxies.
The spread in the values of [N/$\alpha$] observed in DLAs
is larger than that observed in
our Galaxy for stars in the same range of metallicity and in dwarf galaxies,
and can be attributed to different formation epochs or chemical evolution histories for these systems.
The high metallicity ([$\alpha$/H] $>-1.3$) DLAs
exhibit values similar to those of dwarf galaxies
giving support to the suggestion that DLAs
could be the progenitors of dwarf galaxies.
At lower metallicities, however, the [N/$\alpha$] decreases,
and there are DLAs with [N/$\alpha$] values below that of any dwarf galaxy,
which could be very young objects, for which there was no time for 
injection of N into the ISM. 

Recently, Prochaska et al. (2002b) have claimed that 
the observed [N/$\alpha$] ratios in DLAs
present a bimodal distribution in the [N/$\alpha$] vs. [$\alpha$/H] plot,
with a small sub-sample exhibiting significantly lower N/$\alpha$
([N/$\alpha$]$\sim -1.5$) than the majority of systems.
The authours consider that a way of 
naturally reproducing a bimodal distribution of [N/$\alpha$] values
is a top heavy IMF, thus reducing 
the nitrogen contribution of intermediate mass stars (IMS).
The bimodality would severely challenge other explanations:
the low -DLAs [N/$\alpha$] are observed during a transient period
occurring within 250~Myr of a star burst 
and before the IMS have released their nitrogen into the ISM;
N yields of IMS significantly decline with decreasing metallicity;
the low [N/$\alpha$] reflects the nucleosynthesis of Population~III stars.
Notice that a test of the top heavy IMF explanation
would be that it predicts a high [$\alpha$/Fe] ratio.

However, in our sample, the bimodality
in the [N/$\alpha$] vs. [$\alpha$/H] plot is less apparent.
Notice that we consider O, S, and Si$_{cor}$ as the $\alpha$-elements,
while Prochaska et al. (2002b) use Si as a proxy of $\alpha$-elements
and do not consider dust-correction.
In our sample, there are three systems 
with determinations of N and $\alpha$ abundances, which have low
($\sim -1.5$) [N/$\alpha$].
Only one system, the DLA at $z=$2.843 toward Q1946+7658
which happens to be that with the lowest metallicity in Figure 3, 
is isolated from the rest of the systems.
The other two low [N/$\alpha$]  DLAs seem to be contiguous to
the DLAs with higher [N/$\alpha$].

Independently from a segregation of a low [N/$\alpha$] sub-sample,
to be confirmed by further observations,
there seems to be a lower bound envelope
in the [N/$\alpha$] vs. [$\alpha$/H] plot,
extending in an arc
from ([$\alpha$/H],[N/$\alpha$]) $\sim(-1.5,-1.5)$ to $\sim(-0.7,-0.9)$.
The two low [N/$\alpha$]  DLAs with [$\alpha$/H]$\sim -1.4$
are located along this envelope, 
forming a sequence with higher [N/$\alpha$] DLAs.
This feature of the [N/$\alpha$] vs. [$\alpha$/H] distribution
is interesting from the point of view of the models discussed in this work,
since it corresponds to a maximum of the specific star formation rate $\nu$
(i.e. the inverse of the star formation time scale),
which is a central ingredient of the chemical evolution process.
Also the upper limits on [N/$\alpha$] are consistent 
both with non-bimodality and an upper limit for $\nu$.

\subsection{Estimating the dust depletion for DLAs without [Zn/Fe] ratios} 

\begin{figure}
\centerline{
\psfig{figure=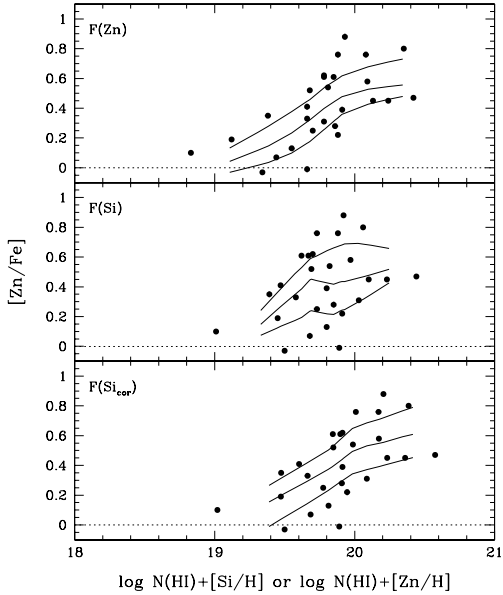,width=8.5cm,angle=0}}
\caption{
The trend of [Zn/Fe] with the metal column density F(X) for
X$=$ Zn, Si, and Si$_{cor}$.
The dotted line represents the solar value. 
The thick solid lines are the statistical fits to the DLA data.
}
\end{figure}

\begin{figure}
\centerline{
\psfig{figure=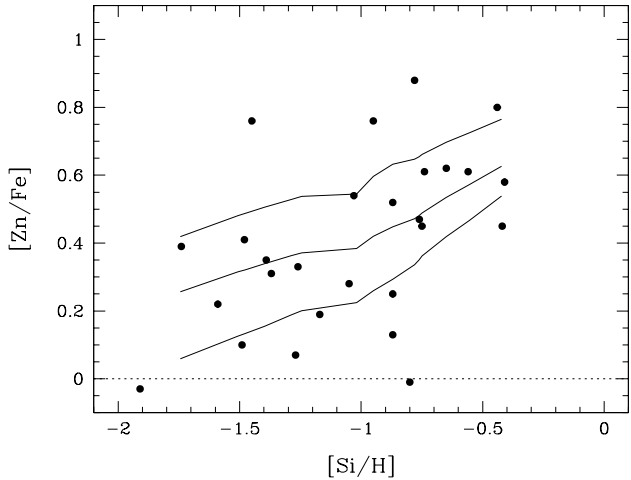,width=8.5cm,angle=0}}
\caption{
The trend of [Zn/Fe] with [Si/H] for the DLA
sample with  Zn, Fe and Si measurements.
The dotted line represents the solar value.
The thick solid lines are the statistical fits to the DLA data.
}
\end{figure}

As any nucleosynthetic contribution to the [Zn/Fe] ratio is small,
not reasonably exceeding $0.3$ dex, the  [Zn/Fe] enhancement
is a reliable indicator of dust depletion.
This is confirmed in Figure 4 
from the good correlation between the
observed [Zn/Fe] ratio and the  
metal column density F(Zn)=log N(HI)+[Zn/H],
which is used in the definition of ``dust filter" by Prantzos \& Boissier (2000),
based on Boiss\'e et al. (1998).
The trend of increasing [Zn/Fe] with F(Zn)
seems to flat beyond F(Zn) somewhat below  F(Zn)=20.0.
One could wonder if obscuration effects due the higher column density
of metals would not prevent the highest depleted media from been seen.
It is interesting to note that this apparent break in the trend
occurs essentially at the same value of F(Zn)=19.8,
beyond which Hou et al. (2001) find evidence for dust depletion in DLAs.

If we consider the equivalent definition of F(Zn) for the silicon,
F(Si)=log N(HI)+[Si/H], the correlation is weaker.
The reason for that, is that the weakly depleted zinc
allows a much more reliable determination of the total
metal column density than the Si, which is mildly depleted.
However, if we use the dust corrected Si for the individul DLAs,
we obtain a much more clear trend of [Zn/Fe] with F(Si$_{cor}$). 
The correlation [Zn/Fe]-F(Si$_{cor}$) 
allow us to be more confident about the dust correction method used in this paper.
In addition, the trend of increasing [Zn/Fe] with F(Si$_{cor}$)
flats for F(Si$_{cor}$)$\ga 20$,
which is the same behaviour exhibited by F(Zn).
However, in this case, the break seems to occur
at a value of F(Si$_{cor}$) that is slightly higher 
that value of the break of F(Zn). 
This would be expected if there is a modest enhancement of the [Si/Fe] ratio.

The considerations above allow us to try to extend the sample of
DLAs with estimated corrected abundances.
In fact, only $\approx 40$\% of the DLAs in our database have been
selected to be applied the dust correction described in the last subsection,
and important pieces of information could be missed.
We would like to estimate intrinsic (dust-corrected) [Fe/Si] ratios
for systems without  Zn measurements.
In the following, we will try to use silicon as an indicator of dust depletion,
or, equivalently, of the Zn/Fe ratio.
Although silicon is a refractory element, its depletion is mild in 
lightly depleted regions of the ISM,
and we have found that the corrections to be applied to Si were never large.
The effects of dust
depletion may hide significant trends with metallicity,
and these effects can be reduced if we use 
a little refractory metallicity indicator.
In fact, dust depletion has spoiled the correlation
between [Zn/Fe] and F(Si), 
as we have already seen from the less clear trend of [Zn/Fe] with F(Si),
when we use the observed (and therefore affected by dust) [Si/H]
instead of the dust-corrected [Si/H].
However, 
as we can see from Figure 5, there is a clear correlation between
[Zn/Fe] and [Si/H], even using the observed [Si/H].

As the [Si/H] metallicity has been revealed to be
a good predictor of the [Zn/Fe] ratio, 
we will extend our sample of DLA with estimated dust depletion
to include those
DLAs which have a value of F(Zn) too small for the Zn lines to be detected.
Hereafter, this sample will be called the ``no Zn sample".
It would allow us to explore a domain with smaller metallicities
on average than the typical [Fe/H]$\ga -2.0$ region defined
by the original sample with Zn and Fe to which we applied the dust
correction procedure.

The [Zn/Fe] ratio, however, at least up to a metallicity [Si/H]$=-0.8$,
varies between zero and values close to 0.8.
Below [Si/H]=-1.0, the dispersion of [Zn/Fe] becomes smaller
and the largest values are bounded by the USMM of the statistical trends,
with the exception of the $z_{abs}=1.776$ DLA toward MC1331+170.
This DLA, however, may be peculiar, since,
in the sample used by Prochaska (2002) 
to investigate abundance variations within DLAs,
it presents the second largest deviation from uniformity.
In order to obtain the no Zn sample, we assume, for the systems
with measurements of Fe and Si, but not of Zn, 
and satisfying [Si/H]$\leq -1.0$, that the [Zn/Fe] ratio varies
between the USMM and the LSMM of the statistical trends shown in Figure 5,
and then apply the dust correction method described in Section 2.2.
We also use some useful upper limits whenever they define
a narrower range of [Zn/Fe] than that delimited by the USMM and the LSMM.
It is the case of the DLAs
toward PKS0528-2505 at $z=$2.141 ([Zn/Fe]$<0.17$),
toward Q0836+11 at $z=$2.465 ([Zn/Fe]$<0.27$),
and toward Q2348-01 at $z=$2.615 ([Zn/Fe]$<0.13$).
The resulting estimates of the dust corrected abundances are shown in Figure 6.
In Figure 6, the polygon encloses the no Zn sample.
In drawing the polygon, we have excluded the DLA at $z=$1.874
toward B2314-409, which has an unusually low observed [Si/Fe] ratio
(Ellison \& Lopez 2001, 2002).
Due to the uncertainties in the dust correction method and
the two-step nature of the process used in building this sample,
assigning dust-corrected abundances has no sense for individual DLAs 
but only for the sample as a whole.
The no Zn sample seems to be an extension toward low metallicities of
the sample with measurements of Zn, Fe and Si.
It is not possible to know if the trend of increasing [Si/Fe] with
decreasing [Fe/H] continues at [Fe/H]$<2$ or if [Si/Fe] reaches a plateau.
For this sample, [Si/Fe]$_{median}=0.204$,
an enhancement only $\sim 0.05$ dex higher than that of the sample
with Zn, Fe and Si measurements.

\begin{figure}
\centerline{
\psfig{figure=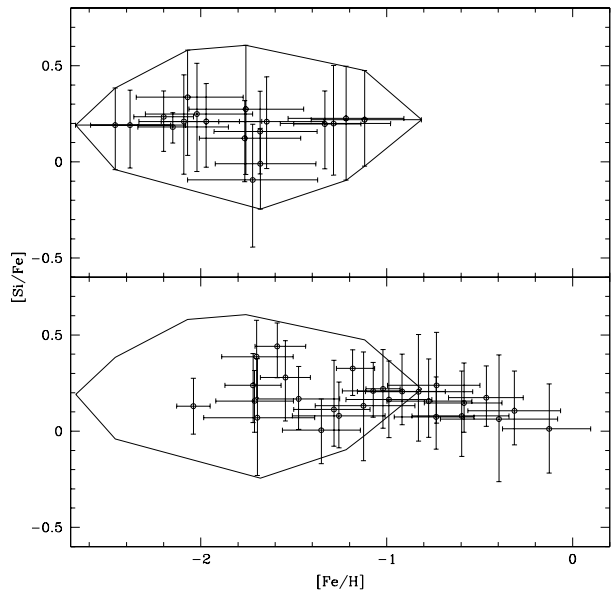,width=8.5cm,angle=0}}
\caption{
Upper panel: The  [Si/Fe] vs. [Fe/H] distribution for the no Zn sample.
Lower panel: Comparison of the Si/Fe] vs. [Fe/H] distributions
of the no Zn sample (represented by polygon enclosing this sample)
with the sample with Zn, Fe and Si measurements.
}
\end{figure}

\section{The chemical evolution models}

In order to study the \ev\ of the metal content in DLAs,
we use two classes of models: 
one-zone \chev\ models and chemodynamical 
models. The one-zone \chev\ models are used to test the disk 
or dwarf galaxy scenario for DLAs,
whereas the chemodynamical model checks for the dwarf galaxy scenario.

For each model class, a \sfor\ law is specified.
The stars formed follow a Salpeter IMF between 0.1 and 100 \msun.
Chemical \ev\ occurs as the stars formed out of the ISM
evolve and eject gas back into the ISM.
The stars are assumed to die either as supernovae (SNe) or as planetary nebulae,
when instantaneous ejection of mass, metals and energy occurs.
The evolution of the abundances of He, C, N, O, Mg, Si, S and Fe
is calculated by solving the basic equations of chemical evolution
(see Matteucci \& Tornamb\`e 1987).
We do not assume instantaneous recycling approximation for the chemical
enrichment, but instead we take into account
the delays for gas restoring from the stars
due to the main-sequence lifetimes.
Instantaneous mixing with the ISM is assumed for the stellar ejecta..
The models start with an entirely  gaseous proto\gal\ or infalling gas
with primordial chemical abundances ($Y=0.24$, $Z=0$).

In this work, we use metallicity dependent yields 
for SNe II, SNe Ia, and IMS.
For SNe II,
we use the metallicity dependent yields of Woosley $\&$ Weaver (1995), 
which are given for stars 
of mass M = 12, 13, 15, 18, 20, 22, 25, 30, 35 and 40 \msun\ and metallicities
Z/Z$_{\odot}$ = 0, 10$^{-4}$, 10$^{-2}$, 10$^{-1}$ and 1. 
Yields of SNIa resulting from Chandrasekhar mass white dwarfs,
are taken from Iwamoto et al. (1999): 
their models W7  (progenitor star of initial metallicity Z=Z$_{\odot}$)
and W70 (initial metallicity Z=0).
The yields for IMS ($0.8 - 8$ \msun),
with initial Z=0.001, 0.004, 0.008, 0.02 and 0.4,
are from van den Hoek \& Groenewegen (1997) (their variable $\eta_{AGB}$ case).
For more details of the nucleosynthesis prescriptions, see 
Fria\c ca \& Terlevich (1998, hereafter FT98).

Our choice of the mass (or surface mass density) and physical size 
of the model galaxies was
based on the scenario in which the amount of gas in DLAs
constitute a reservoir for the formation of present day galaxies,
in view of the coincidence of the amount of mass in DLAs
and the baryonic stellar mass in galaxies today 
(Storrie-Lombardi \& Wolfe 2000).
Therefore, the column density of the models should be not far
from the baryonic column density of the present day
galaxies being modelled.
At the same time, the H I column densities predicted by
all the models satisfy both our DLA definition, log N(HI)$\ge 20.0$,
and the high column density cut-off 
log N(HI)$\la 22$, observed in DLAs, and explained by
Schaye (2001) as due the conversion of atomic hydrogen to molecular hydrogen
in clouds with higher total hydrogen (HI+H$_2$) column density.
Therefore, in comparing the observations with the predictions of the models,
we will consider not only the abundances and abundances ratios,
but also the H I column densities.

\subsection{The one-zone chemical evolution models}

Two types of one-zone \chev\ models are used to explore
the two most popular hypothesis for the nature of DLAs: 
disks (Prochaska $\&$ Wolfe 1997),
and dwarf galaxies (Centuri\'on et al. 2000, Molaro et al. 2001). 
One-zone models with galactic winds are used for dwarf galaxies,
and multi-zone chemical evolution models with infall are used to 
represent disks.

\subsubsection{The dwarf galaxy model}

The dwarf galaxy model is the classical model
with galactic winds (Matteucci 1994),
first proposed by Larson (1974).
In this model a wind is established at a time $t_w$ when the 
thermal energy of the gas becomes higher than the potential 
energy, $E_{th,SN}\;>\;E_{b,gas}$. 
The wind instantly sweeps out all the gas of the galaxy
and the star formation ceases. 
Later, processes of stellar mass loss restore the ISM.

Most of the models represent a galaxy with 
initial gas mass $M_G=10^9$ $\msun$,
but there is also a model with $M_G=10^7$ \msun.
The galaxies are inside a dark halo with a mass 7.5 times $M_G$. 
The star formation rate (SFR) is proportional to the present mass 
of gas and characterised by the specific star formation 
rate $\nu$ (the inverse of the star formation time scale).
The fact that these models are simple,
parameterized only in terms of $\nu$ 
makes them useful for gauging the effect of the star formation rate
in the \chev\ of DLAs.
We run models for several values of $\nu$
ranging from 0.1 to 19 Gyr$^{-1}$. 
$\nu=19$ Gyr$^{-1}$ follows from assuming $M_G=10^9$ \msun\ in the
relation $\nu= 8.6 (M_G/10^{12} \msun)^{-0.115}$ used in
Matteucci \& Tornamb\'e (1987) and later works.
The values of $\nu \leq 3$ Gyr$^{-1}$ reflect the 
typically low SFR of DLAs, as inferred from observations.

Interesting limits on the SFR of DLAs are given by
imaging observations of the $z_{abs}=3.386$ DLA toward Q0201+1120
(Ellison et al. 2001a).
There are only two faints sources near the quasar with photometric
redshifts consistent with the redshift of the DLA, the LBGs G2 and oM6
separated from the QSO by 4 and 2.9 arcsec.
Spectroscopy of oM6 (${\cal R}$=23.97 mag) revealed 
a prominent Lyman $\alpha$ emission line
at $z_{em}=3.645$. Therefore, oM6 is not the DLA.
Rather, it is associated to the QSO
(the difference in redshift
between the oM6 and Q0201+1120 is only +465 \kms).
G2,  the remaining candidate for DLA,
is so faint (${\cal R}$=25.3) that there is no spectroscopy available.
However, assuming that G2 is at the redshift of the DLA,
we can estimate its SFR from its ${\cal R}$ magnitude,
because, at $z \sim 3$, 
the ${\cal R}$ filter samples the rest-frame far-UV spectrum produced
by recently formed O and B stars.
Assuming that G2 has a typical DLA metallicity of [Fe/H]$=-1.5$
and that is undergoing continuous \sfor\ for $10^8$ yr,
we obtain, following Fria\c ca \& Terlevich (1999),
a SFR of $6.2$ \msun\ yr$^{-1}$.
This corresponds to a model with $M_G=10^9$ \msun\ and  $\nu=6$ Gyr$^{-1}$
(or $M_G=2\times 10^9$ \msun\ and  $\nu=3$ Gyr$^{-1}$).
However, if G2 is not the DLA, then the absorber must be fainter than
${\cal R}$ $\approx 26.0$, and the corresponding SFR
$\la 3$ \msun\ yr$^{-1}$, or $\nu < 3$ Gyr$^{-1}$ for $M_G=10^9$ \msun.

\subsubsection{The disk model}

The disk model belongs to the class of the models for the Galaxy
with one infall episode of pristine gas.
Following Matteucci \& Fran\c cois (1989),
the SFR depends on the galactocentric radius $r$
through the total mass surface density $\sigma(r,t)$
and the gas fraction $G(r,t)$ as:

\begin{equation}
\Psi(r,t)=\tilde{\nu} \; \biggl[{\sigma(r,t) \over \tilde\sigma(\tilde r,t)}
\biggl]^{2x_{SF}} \; \biggl[{\sigma(r,t_G) \over \sigma(r,t)}\biggl]^{x_{SF}}
\; G(r,t)^{(x_{SF}+1)} \;. \\
\end{equation}

\noindent
The normalisation $\tilde\nu$ of the 
SFR law is set at the solar position ($\tilde r=8$ kpc) and 
at a galaxy age $t_G=13$ Gyr 
(corresponding to a galaxy formation epoch  $z_{GF}=10$).
We assume $\tilde\nu=0.5$ Gyr$^{-1}$
and $x_{SF}=1/2$ in the adopted SFR law,
in which the specific \sfor\ rate depends on a power of the gas density
$\nu \propto \rho^{x_{SF}}$.
The infall rate into the disk
is assumed to decrease with time as $e^{-t/\tau _D}$,
with the infall time scale $\tau _D$ given by:

$\tau _D$ (r) = 1 + 7/6(r-2),  for 2 $\leq $  r  $\leq $ 8 kpc 
 
$\tau _D$ (r) = 8 + 0.5(r-8),  for r $>$ 8 kpc 

\noindent
In this way, $\tau _D=8$ and 1 Gyr at the solar position and at $r=2$ kpc,
respectively (Chiappini, Matteucci \& Graton 1997),
and $\tau _D=13$ Gyr at $r=18$ kpc.
An exponential surface mass density profile is adopted for the disk,
$\sigma \propto e^{-r/r_{G}}$,  
with $r_G$ = 2.6 kpc (Boissier $\&$ Prantzos 2000),
normalized to $\sigma(\tilde r,t_G)=50$ \msun\ pc$^{-2}$ (Chiappini et al. 1997).
We do not consider any threshold for \sfor, differently from,
for instance, Chiappini et al. (1997),
who adopt the threshold $\Sigma_{\rm th}=7$ \msun\ pc$^{-2}$.
Adopting a threshold  would impact on the 
outer regions of the disk, where it would induce a
starburst-like star formation regime (Calura et al. 2002).

The disk model, the column density of HI implied 
by the present day total surface
mass density is $4.7\times 10^{21}$ \cc\ at the solar radius
and $1.0\times 10^{20}$ \cc\ at $r=18$ kpc,
which is consistent with the range of N(HI)values in DLAs.
During the evolution of the disk
(not in the very early phases, i.e. $t < \tau_D(r)$,
when there is still very little gas in the disk),
the value of the column density will
vary by a factor of few around the values above.

\subsection{The chemodynamical model}

We investigate the scenario in which the DLAs are dwarf galaxies
with the chemodynamical model of FT98, in which a single massive dark halo
hosts baryonic gas that will fall toward the centre of dark halo
and will subsequently form stars.
The code combines a multi-zone chemical evolution solver and a 1D 
hydrodynamical code. 
The system, assumed to be spherical, is 
subdivided in several spherical zones and the 
hydrodynamical evolution of its ISM is 
calculated. The equations of chemical evolution for each 
zone are then solved taking into account the gas flow,
and the \ev\ of the chemical abundances is obtained.
A total of $\approx$100 star generations are stored during 13 Gyr 
for chemical evolution calculations. 
We assume that at a given radius $r$ and and the time $t$,
the specific \sfor\ rate $\tilde \nu(r,t)$
follows a power-law function of gas density ($\rho$):
$\tilde \nu(r,t)=\nu (\rho/\rho_0)^{1/2}$,
where $\rho_0$ is the initial average gas density inside the core radius
of the dark halo ($r_h$), and $\nu$ is the normalization of the \sfor\ law.
We include inhibition of star formation for expanding gas ($\nabla.u>0$)
or when the density is too low, and, therefore, the cooling is inefficient
(i.e., for a cooling time $t_{coo}=(3/2)k_B T/\mu m_H \Lambda(T) \rho$
longer than the dynamical time $t_{dyn}=(3\pi/16\,G\,\rho)^{1/2}$)
by multiplying $\tilde \nu$ as defined above by the inhibition factors
$(1+t_{dyn}\,{\rm max}(0,\nabla.u))^{-1}$ and $(1+t_{coo}/t_{dyn})^{-1}$.
A characteristic of these models is that several episodes of inflow
and outflow occur simultaneously ar different radii.
The chemodynamical model for spheroids was used to investigate 
the relation between young elliptical galaxies and QSO activity
(FT98),
the absence of passively evolving elliptical galaxies in deep surveys
(Jimenez et al. 1999),
LBGs (Fria\c ca $\&$ Terlevich 1999),
Blue Core Spheroids (Fria\c ca $\&$ Terlevich 2001),
and the coupled spheroid and black hole formation (Archibald et al. 2002).

We have extended the model family used in previous works 
(FT98, Fria\c ca $\&$ Terlevich 1999, 2001) 
toward lower masses to study dwarf spheroids.
The model dwarf galaxy has an initial gas mass $M_G=10^9$ \msun, and
the normalization of the SFR was set $\nu = 3$ or 1 Gyr$^{-1}$,
rather than 10  Gyr$^{-1}$ as for luminous ellipticals
(in order to reproduce the suprasolar [Mg/Fe] ratio of giant ellipticals).
The scaling laws used in FT98, 
which are valid for giant or intermediate ellipticals,
were changed to forms more appropriated for dwarf galaxies,
using as reference the ``bright dwarf ellipticals"
defined in Bender, Burstein \& Faber (1993):
1) $r_h=1$ kpc was kept as in the 10$^{10}$ \msun\ model, since dwarf galaxies
have a larger radius with respect to their masses than bright ellipticals;
2) $r_{tidal}$/ $r_h$ = 14 instead of 28, used for
luminous/intermediate ellipticals;
3) the Faber-Jackson relation was
redefined in order to fit the pair NGC205-NGC3605.
In FT98, the stellar population of the model \gal\ follows a Faber-Jackson relation,
$\sigma_*=200(L_B/L_B^*)^{1/4}$ km~s$^{-1}$,
with $L_B$ related to $M_G$ through $[M/L_B]=10$,
the mass-light ratio typical of an $L^*$ galaxy,
or, equivalently, 
$\sigma_*=150.5(M/10^{11}\,\msun)^{1/4}$ km~s$^{-1}$.
Here, we take the dynamical masses and velocity dispersions
determined by Bender at al. (1993) for NGC205 and NGC3605 to obtain
$\sigma_*=188.3(M/10^{11}\,\msun)^{0.363}$ km~s$^{-1}$.
The initial HI column density of the model is
N(HI)$=1.65\times 10^{21}$ \cc\ through the centre of the galaxy.

\section{Investigation of DLAs with chemical evolution models}

\subsection{One-zone model for dwarf galaxies}

\begin{figure}
\centerline{
\psfig{figure=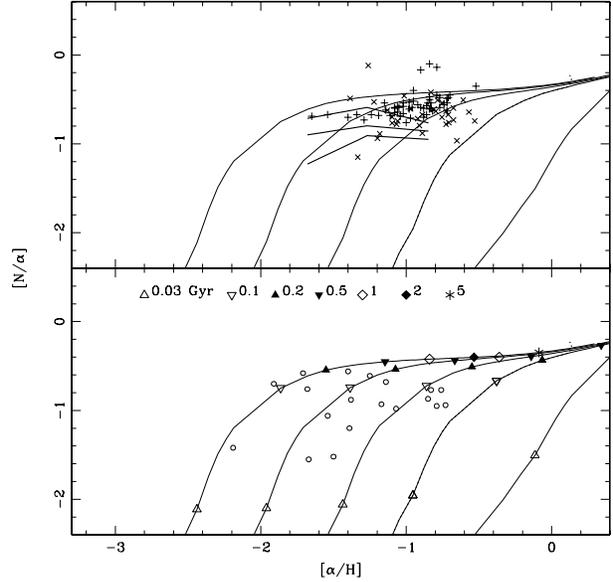,width=8.5cm,angle=0}}
\caption{[N/$\alpha$] vs. [$\alpha$/H] observed in DLAs and dwarf 
galaxies compared with one-zone model predictions for dwarf galaxies.
The solid lines represent the grid of models with $\nu$ = 0.1, 0.3, 
1, 3, 19 $Gyr^{-1}$ (left to right), 
and the dotted line (barely seen) a model with
$\nu=3.0$ Gyr$^{-1}$ and $M=10^{7}$ \msun. 
Several evolutionary times are indicated by the symbols on the model lines.
The lower panel exhibits the [N/$\alpha$] vs. [$\alpha$/H] 
data for DLAs (open circles),
and upper panel the data for dwarf galaxies
from van Zee et al. (1997a,b) (crosses)
and Izotov $\&$ Thuan (1999) (plus signs).
The $\alpha$-element abundance is given
by O for the model predictions and for the dwarf galaxy data, and
for the DLA data, by O or S, or, when not available, by dust-corrected Si.
The thick solid lines in the upper panel are the statistical fits
to the DLA data.
}
\end{figure}

The predictions for [N/$\alpha$]
of the classic wind model for dwarf galaxies are compared 
to the observed ratios both in DLAs and dwarf galaxies
in Figure 7. 
The observed [N/$\alpha$] in DLAs shows a larger dispersion
than that found for Galactic stars in the same range of metallicity
and for dwarf galaxies. 
The different values of [N/$\alpha$]
for a given [$\alpha$/H] can be explained by the different
time scales for production of N and $\alpha$ elements,
as one can see in the sequence of models
with star formation ranging from almost quiescent
($\nu=0.1$ Gyr$^{-1}$) to violent ($\nu=19$ Gyr$^{-1}$). 
Most dwarf galaxies seem to require a specific star 
formation $\nu$ between 0.3 and 3 Gyr$^{-1}$,  
while the typical $\nu$ for DLAs seems to be somewhat lower,
with a significant number of DLAs with $\nu \approx 0.1$ Gyr$^{-1}$.
For both populations,  $\nu \approx 3$ Gyr$^{-1}$ seems to be an upper limit
for the efficiency of the SFR.

In our models, when the galactic wind occurs, the metallicity of the system
is too high for a DLA.
The time of the onset of the is galactic wind is $t_w=0.16$, 1.47 and 5.66 Gyr,
for $\nu=19$, 3 and 1 Gyr$^{-1}$.
The models with $\nu=0.3$ and 0.1 Gyr$^{-1}$ do not develop galactic wind
until $t=13$ Gyr.
The effect of the galactic wind is illustrated by the model with
$M_G=10^7$ \msun.
When the galactic wind occurs at 0.3 Gyr, the \sfor\ stops, and,
consequently, the O release by SNII; however, the IMS keep ejecting
N to the ISM.
This is the reason for the upturn of the
$M_G=10^7$ \msun\ model evolutionary track at [$\alpha$/H]$\approx 0.1$,
away from the $M_G=10^9$ \msun\ model path.

\begin{figure}
\centerline{
\psfig{figure=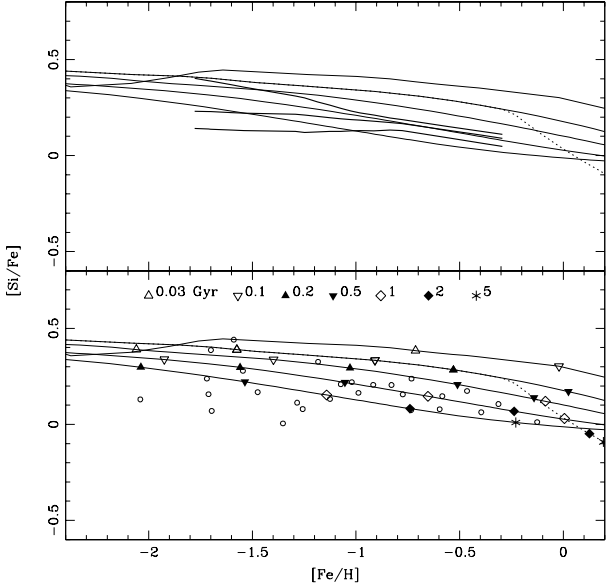,width=8.5cm,angle=0}}
\caption{[Si/Fe] vs. [Fe/H] observed in DLAs (open circles) compared with 
one-zone models predictions for dwarf galaxies. 
The data points represent the dust-corrected abundances. 
The solid lines represent the grid of models with $\nu$ = 0.1, 0.3, 
1, 3, 19 $Gyr^{-1}$ (bottom to top),
and the dotted line a model with $\nu=3.0$ Gyr$^{-1}$ and $M=10^{7}$ \msun. 
The thick solid lines in the upper panel are the statistical fits.
}

\end{figure}

The predictions for [Si/Fe] of dwarf galaxy models with 
different values of $\nu$ are compared to the observed values in Figure 8.
Most systems have [Si/Fe] ratios below the predictions of any model. 
The classic wind model fails in explaining a large number of systems
with near-solar [Si/Fe] ratios, unless we assume unrealistic,
extremely low values of $\nu$ ($< 0.1$ Gyr$^{-1}$).

\subsection{One-zone model for disk galaxies}

\begin{figure}
\centerline{
\psfig{figure=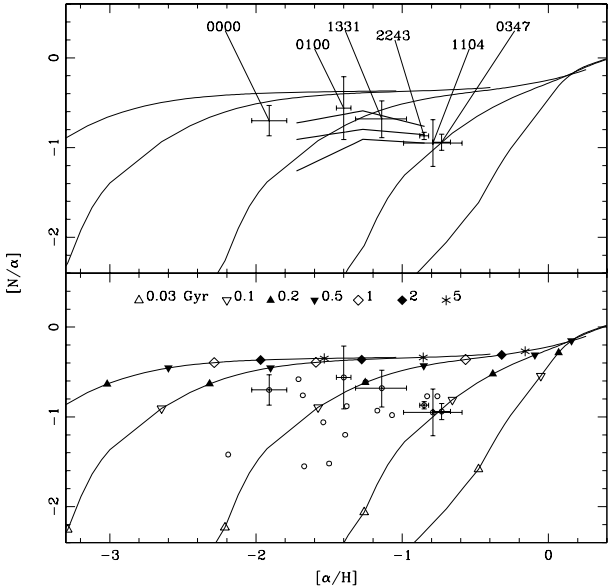,width=8.5cm,angle=0}}
\caption{[N/$\alpha$] vs. [$\alpha$/H] observed in DLAs (open circles) 
compared with 
predictions from models of disk galaxies. 
The solid lines are the predictions of the disk model
at r = 2, 4, 8, 14, 18 kpc (right  to left).
The thick solid lines in the upper panel are the statistical fits.
The points with error bars refer to DLAs 
with both [N/$\alpha$] and [Si/Fe]$_{cor}$ determinations, 
and are identified by the right ascension of the background QSO.
}
\end{figure}

The disk models predictions are compared to
the observed values in DLAs (Figures 9 and 10).
There is a great dispersion for the [N/$\alpha$] ratio,
but the observed [N/$\alpha$] 
ratios are confined to the region between r = 4 and 14 kpc, 
with a significant number of systems around r = 8 kpc.
The statistical trend lines seem to confirm the behaviour
predicted by the models: the 
ratio increases with [$\alpha$/H] rapidly in low metallicities
and then becomes flatter.
The plateau at higher metallicities is due to primary N production by IMS.
The comparison between the observed [N/$\alpha$] with
the disk model predictions indicates star formation time scales
in DLAs similar to those found in the solar neighbourhood.
The lower bound envelope in the [N/$\alpha$] vs. [$\alpha$/H]
distribution of the DLA data is nearly traced by the
evolutionary track at $r=4$ kpc.

With respect to the [Si/Fe] predictions from the disk model,
the region $r \approx 8 - 18$ kpc accounts for most of the systems.
There are, however, a few systems with higher [Si/Fe] ratios
implying a faster SFR than that found in outer regions of disks.
The models also present the same increasing trend
of [Si/Fe] with decreasing [Fe/H] of the data.
Since external regions ($r \ga 8$ kpc) are favored as sites for DLAs,
long star formation time scales are typical for these systems.
A long and continuous star formation allows the iron peak group 
elements to be produced by SN Ia, giving rise to low values of [Si/Fe].
A slow \sfor\ was also inferred from the analysis of the [N/$\alpha$] ratios,
but with the difference that more central regions
($r \approx 4 - 14$ kpc) are required by the data.
Although the [Si/Fe] ratio apparently has a smaller dispersion
than the [N/$\alpha$] ratio, 
a number of systems have values of [Si/Fe] too low or too high
to be explained as arising in one characteristic portion of a disk.
In addition, there are a few systems with [Si/Fe] lower than that expected
even in the outermost regions of a  disk,
and, for a number of DLAs, their low [Si/Fe] and high [Fe/H] could
be achieved in the outer disk but at times later ($t \sim 5$ Gyr or later)
than those allowed by the age of Universe at the DLA redshift.

\begin{figure} 
\centerline{ 
\psfig{figure=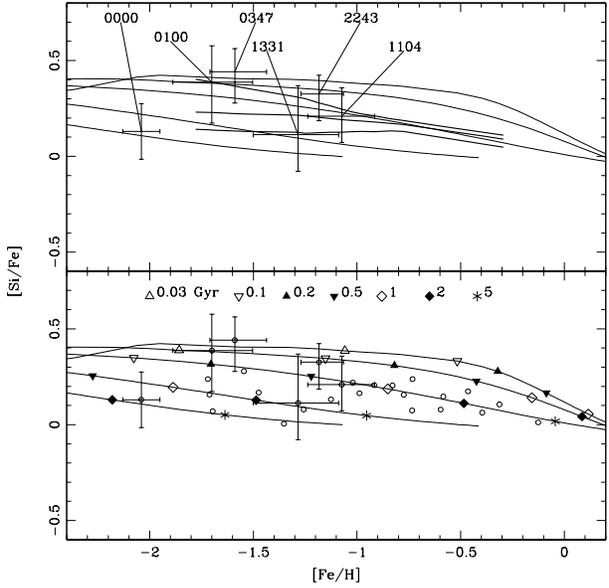,width=8.5cm,angle=0}} 
\caption{[Si/Fe] vs. [Fe/H] observed in DLAs (open circles) compared with 
chemical evolution models. 
The points represent the dust-corrected abundances. 
The solid lines are the predictions of the disk model
at r = 2, 4, 8, 14, 18 kpc (top to bottom). 
The thick solid lines in the upper panel are the statistical fits.
The points with error bars have the same meaning as in Figure 9.
} 
\end{figure} 

Figure 11 presents the evolution the metallicity
as a function of the H I column density predicted by
the disk model for the radii r = 2, 4, 8, 14, 18 kpc.
This model for the Galactic disk predicts, at
the outermost radius, metal column densities 
([Si/H]+log N(H I)) that are too low at all times.
The innermost radius that match the metallicity-N(H I)
constraints is $r=14$ kpc, but only at late times,
$t>3 $ Gyr, thus limiting the applicability of this model
to low redshift objects only.
At the solar radius and $r=4$ kpc, the predictions match well the observations.
The $r=2$ kpc zone also satisfy the constraints in Figure 11,
but only at very early times ($t\la 5\times 10^7$ yr).

\begin{figure} 
\centerline{ 
\psfig{figure=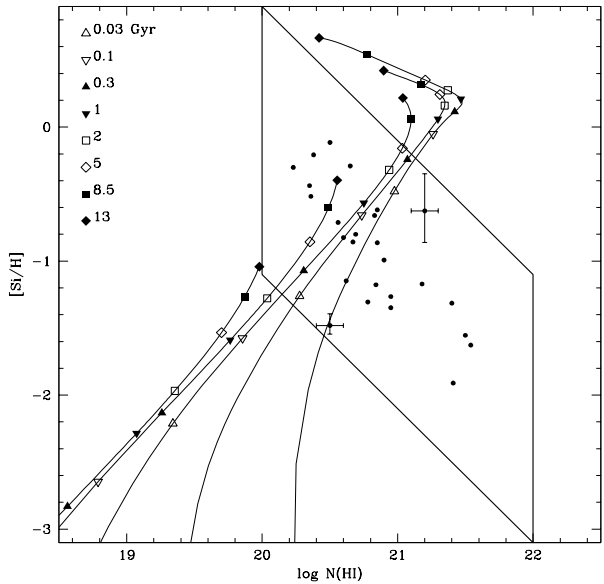,width=8.5cm,angle=0}} 
\caption{
Metallicity vs. H I column density
observed in DLAs compared with 
chemical evolution models. 
The metallicity is given by [Si/H]$_{cor}$ for the DLAs
(filled circles).
The solid lines are the predictions for disk models
at r = 2, 4, 8, 14, 18 kpc (left to right). 
The thick line contour encloses the data according to the
general constraints on HI column density for DLAs,
$20 \leq \log N(HI) \leq 22$,
and the limits on the metal column density of our sample,
F([Si/H]$_{cor}$)=[Si/H]$_{cor}+\log N(HI)\leq 20.9$ (upper diagonal)
and  F([Si/H]$_{cor}$)$\geq 18.9$ (lower diagonal).
For the sake of clarity, the errors are shown only for the two data points with
the lowest and highest F([Si/H]$_{cor}$).
} 
\end{figure}

\subsection{The chemodynamical model for dwarf spheroids}

The predictions for the [N/$\alpha$] ratio of
the chemodynamical model for dwarf spheroids
are compared to the observed ratios in Figures 12 and 13. 
One characteristic of the model is the occurrence of several short
starbursts over the galaxy, which lead to rapid excursions
of the evolutionary tracks in the [N/$\alpha$] vs. [$\alpha$/H] plane.
Most of the systems are young, with ages around 0.1 Gyr or less
(three systems, with [N/$\alpha$]$\sim -1.5$, have indicated ages
of $\sim 0.05$ Gyr).
The DLAs with [N/$\alpha$]$>-0.7$ and [$\alpha$/H]$<-1.0$
could be older systems ($t>0.1$ Gyr),
but the crowding of evolutionary tracks prevents us from
assigning narrow ranges of ages (and of radii) to them. 

The lower bound envelope in the [N/$\alpha$] vs. [$\alpha$/H]
distribution of DLAs,
which was reproduced by the one-zone dwarf galaxy model
and by the disk model, representing an upper limit on $\nu$,
is even more apparent in the chemodynamical model with $\nu=3$ Gyr$^{-1}$,
and corresponds to the radius $r\approx 1$ kpc.

\begin{figure}
\begin{center}
\centerline{
\psfig{figure=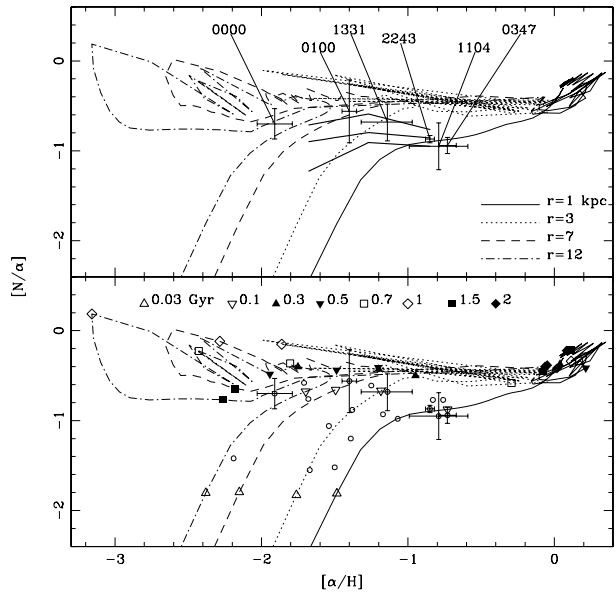,width=8.5cm,angle=0}}
\caption{[N/$\alpha$] vs. [$\alpha$/H] observed in DLAs (open circles)
compared to the predictions of chemodynamical model for dwarf spheroids
with $\nu=3$ Gyr$^{-1}$.
The lines represent the model predictions at several radii.
The thick solid lines in the upper panel are the statistical fits.
The points with error bars have the same meaning as in Figure 9.
}
\end{center}
\end{figure}

\begin{figure}
\begin{center}
\centerline{
\psfig{figure=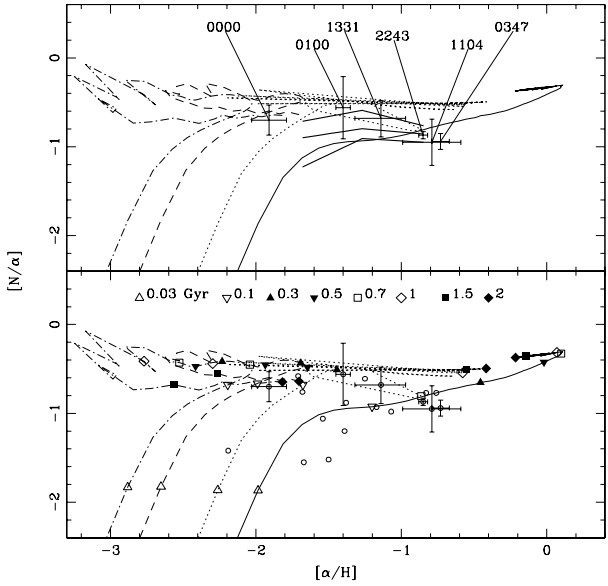,width=8.5cm,angle=0}}
\caption{[N/$\alpha$] vs. [$\alpha$/H] observed in DLAs (open circles)
compared to the predictions of chemodynamical model for dwarf spheroids
with $\nu=1$ Gyr$^{-1}$.
The lines represent the model predictions at several radii.
The meaning of the line styles is the same as in Figure 12.
The thick solid lines in the upper panel are the statistical fits.
The points with error bars have the same meaning as in Figure 9.
}
\end{center}
\end{figure}

The predicted and observed [Si/Fe] ratios in DLAs are shown in Figures 14 and 15.
Nearly all [Si/Fe] values of DLAs are reproduced in the
$r \sim 1-7$ kpc region,
which is characterized by a low to intermediate star
formation rate.
As a consequence of the starbursts, 
at $r\sim 3$ kpc, the [Si/Fe] ratio varies abruptly
allowing the evolutionary track of the model to cover most of the data points.
Multiple starburst also happen at $r\sim 1$ kpc,
although less dramatically seen in the [Si/Fe] vs. [Fe/H] diagram.
The evolutionary tracks of the model show for [Fe/H] $<-2$ a trend of increasing
[Si/Fe]  with decreasing metalicity which may be confirmed or not by the
no Zn sample, with typically lower metallicities.
More reliable abundance measurements in this range of metallicities are
needed in order to establish whether there is discrepancy between the data
and the models. 
Note that a potential discrepancy could be used as a test for the SNe II yields.

\begin{figure}
\begin{center}
\centerline{
\psfig{figure=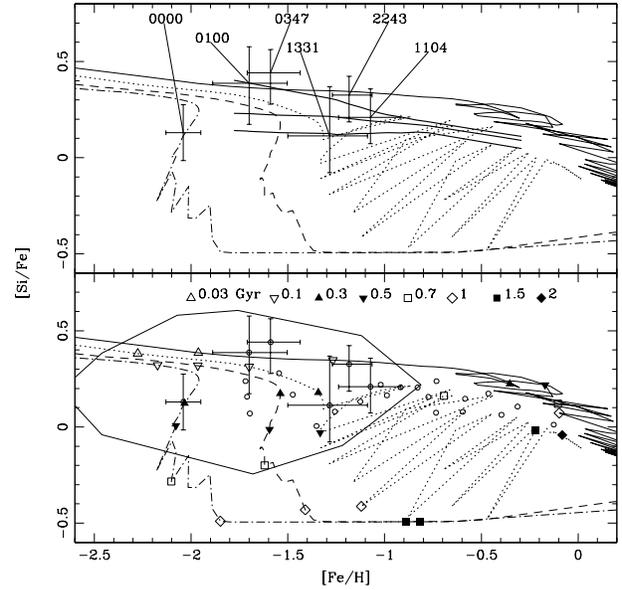,width=8.5cm,angle=0}} 
\caption{[Si/Fe] vs. [Fe/H] observed in DLAs (open circles)
compared to the predictions of the chemodynamical model for dwarf spheroids
with $\nu=3$ Gyr$^{-3}$.
The lines represent the model predictions at several radii.
The meaning of the line styles is the same as in Figure 12.
The thick solid lines in the upper panel are the statistical fits.
The points with error bars have the same meaning as in Figure 9.
The polygon encloses the no Zn sample defined in Section 2.3.
}
\end{center}
\end{figure}

\begin{figure}
\begin{center}
\centerline{
\psfig{figure=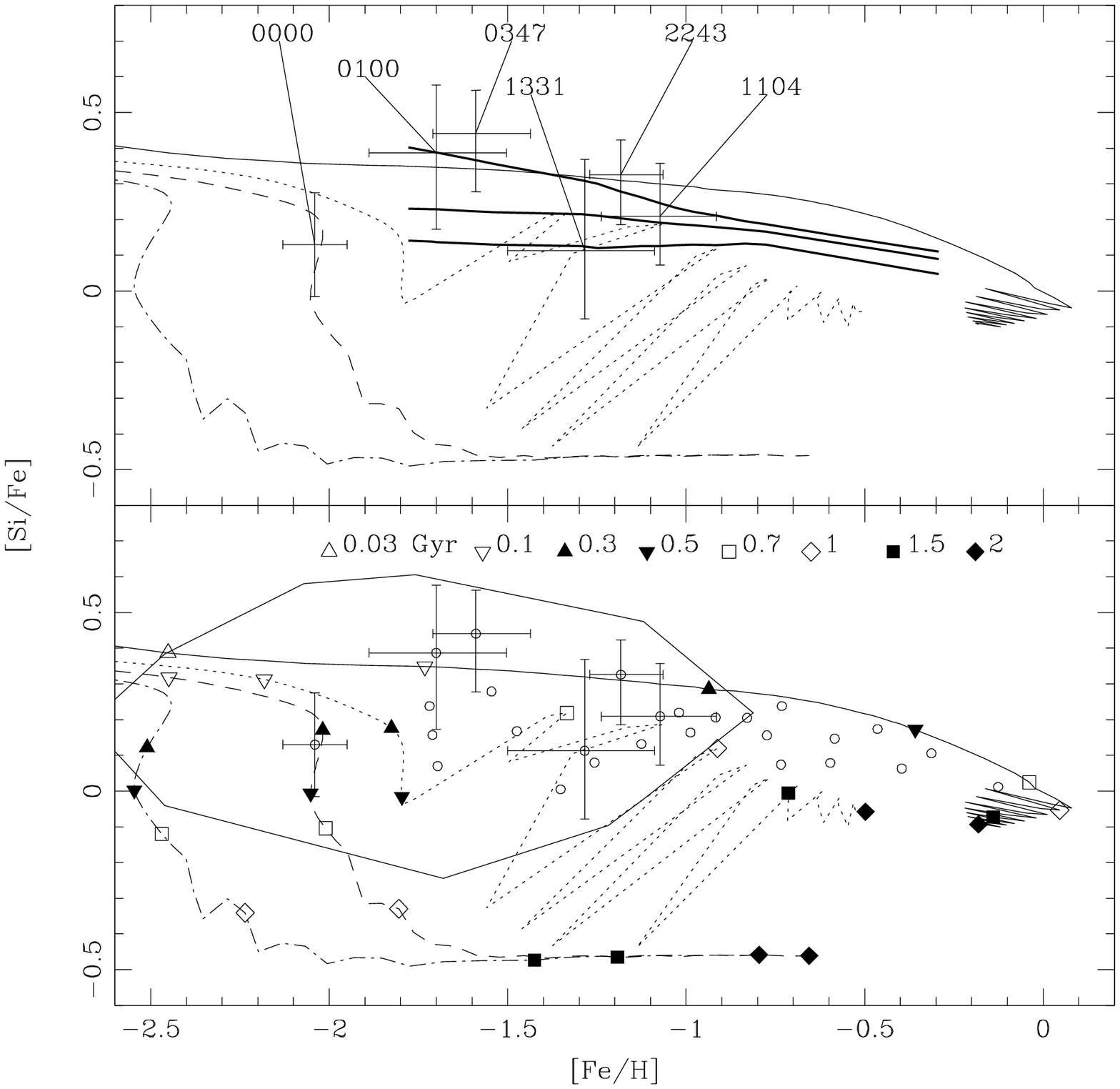,width=8.5cm,angle=0}} 
\caption{[Si/Fe] vs. [Fe/H] observed in DLAs (open circles)
compared to the predictions of the chemodynamical model for dwarf spheroids
with $\nu=1$ Gyr$^{-3}$.
The lines represent the model predictions at several radii.
The meaning of the line styles is the same as in Figure 12.
The thick solid lines in the upper panel are the statistical fits.
The points with error bars have the same meaning as in Figure 9.
The polygon encloses the no Zn sample defined in Section 2.3.
}
\end{center}
\end{figure}

The occurrence of multiple starbursts and the peculiar tracks 
in both diagrams
 is explained by the connection between
star formation and gas flows.
In the dwarf galaxy model, there is a first infall episode
of primordial gas into the dark matter halo potential well
which ends with a starburst.
This first starburst drives an outflow, which 
is very weak and it is reverted into infall by the large outer gas reservoir.
During the outflow (except in its beginning) and the subsequent infall,
the gas density is not high enough to allow \sfor.
Since the infall carries little enriched gas from
the outer parts of the galaxy, there is a decrease in the metallicity
of the gas as the infall goes on.
Eventually the second infall builds up a gas density high enough
to trigger a second starburst.
In the inner regions of the galaxy, the cycle restarts a number of times
giving rise to a series of bursts of \sfor.
In contrast, in the outer ($r \ga 4$ kpc) regions of the galaxy, 
the cycle happens only once,
and the gas succeeds in leaving the galaxy.
Only this region presents a true galactic wind,
which is enriched by SNe Ia.
The plateau at low, constant [Si/Fe] and [Fe/H] increasing with time,
which is reached at $t \approx 1$ Gyr in the outer regions,
marks the onset of the wind driven by SNe Ia.
Therefore, dwarf spheroids contribute to the enrichment of the 
intracluster medium
mainly in iron and not in $\alpha$ elements (see Fria\c ca 2000).
The galactic wind is not a favourable site for DLAs
--- the low densities and high temperatures ($T \sim 10^7$ K)
make the cooling time too long for the development
of thermal instabilities into DLAs
--- which accounts for the absence of DLAs in the galactic wind plateau.

\begin{figure} 
\centerline{ 
\psfig{figure=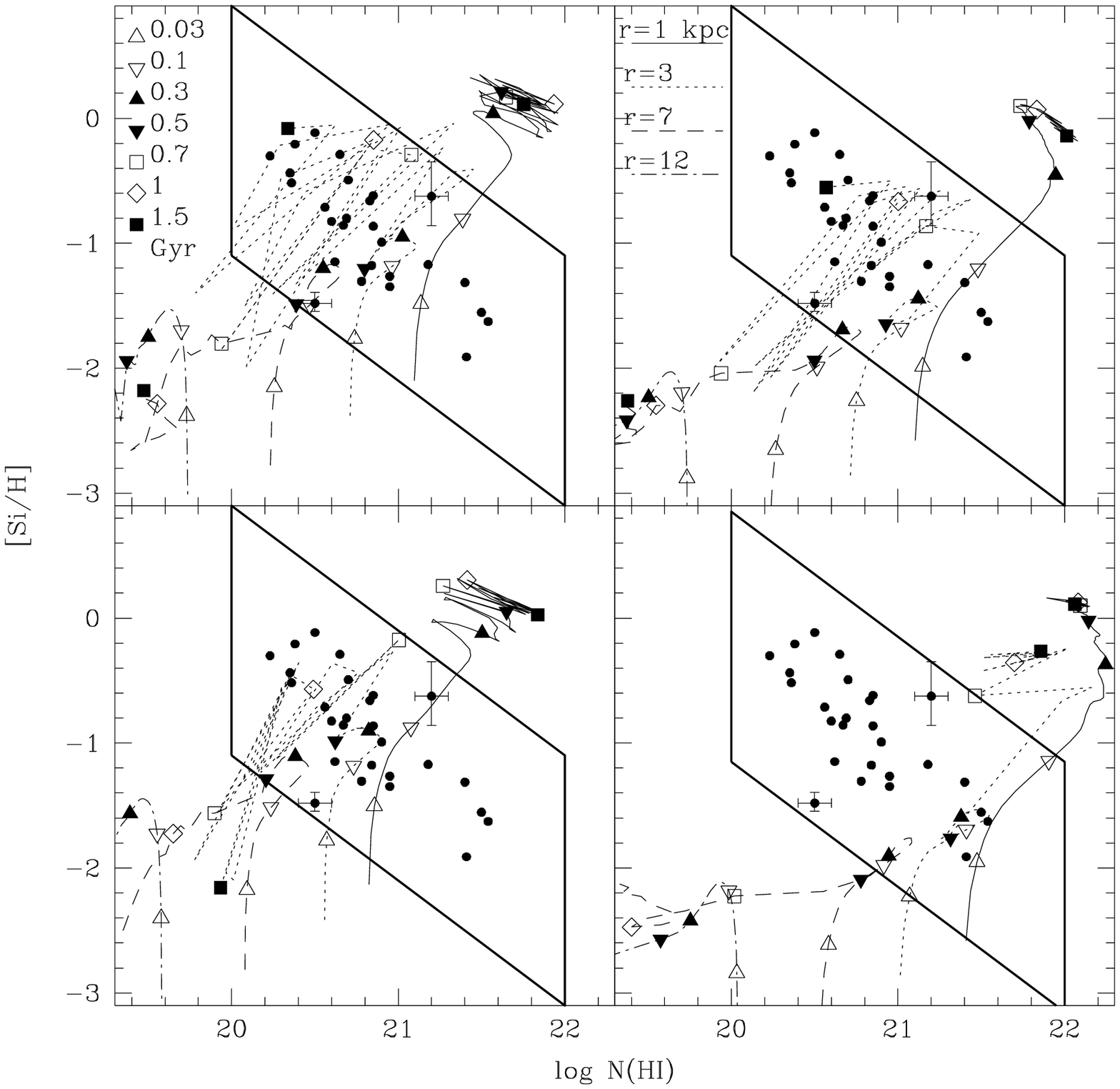,width=8.5cm,angle=0}} 
\caption{
Metallicity vs. H I column density observed in DLAs 
compared to the predictions of the chemodynamical models for dwarf spheroids
with $M_G=10^9$ \msun and $\nu=3$ Gyr$^{-1}$ (upper left panel)
and $\nu=1$ Gyr$^{-1}$ (upper right panel),
with $M_G=5\times 10^8$ \msun\ and $\nu=3$ Gyr$^{-1}$ (lower left panel)
and with $M_G= 2\times 10^9$ \msun\ and $\nu=1$ Gyr$^{-1}$ (lower right panel).
The lines represent the model predictions for several impact parameters.
The meaning of the line styles is the same as in Figure 12.
The metallicity is given by [Si/H]$_{cor}$ for the DLAs (filled circles).
The thick line contour encloses the data according to the
general constraints on HI column density for DLAs, $20 \leq \log N(HI) \leq 22$,
and the limits on the metal column density of our sample,
F([Si/H]$_{cor}$)=[Si/H]$_{cor}+\log N(HI)\leq 20.9$ (upper diagonal)
and  F([Si/H]$_{cor}$)$\geq 18.9$ (lower diagonal).
For the sake of clarity, the errors are shown only for the two data points with
the lowest and highest F([Si/H]$_{cor}$).
} 
\end{figure}

Figure 16 shows 
the joint evolution of the H I column density
and of the metallicity predicted by the chemodynamical model
for several impact parameters.
In this paper, we define as HI gas that with $T\leq 2\times 10^4$ K.
During the evolution of the galaxy, part of the gas is heated mainly by SNe
to temperatures higher than this limit.
The hotter gas is not included in the
calculation of the column density.
At the start of the calculations the \chdyn\ model
exhibits N(HI)$=1.23\times 10^{21}$ \cc, 
through an impact parameter of 1 kpc (i.e. at $r=r_h$).
The initial infall and, after a while, a series of
starbursts are visible through excursions
in the log N(HI) vs. [Si/H] diagram.
The starbursts have a more dramatic effect in the
$r=3$ kpc region.
In this region, the starbusts shift along the strip
defined by the observational limits on the metal column density,
toward higher metallicities and lower values of N(HI).
In the outer regions ($r=7$ and $r=12$ kpc),
after the early infall, the \gw\ prevails and N(HI) is reduced
dramatically.
Note that at $r=12$ kpc, differently from at more central impact parameters,
there is no initial increase of N(HI)
due to the baryonic infall, and N(HI) remains below the lower limit for a DLA.
Rather, the infall accumulates mass only
in the inner regions of the galaxy.
The outskirts of the galaxy ($r\ga 10$ kpc) would host only
sub-DLAs and Lyman Limit Systems (LLSs).
(sub-DLAs are QALs with N(HI)$=10^{19}-10^{20}$ \cc 
--- this range is narrower than that of the original definition of sub-DLAs, 
N(HI)$=10^{19}-2\times 10^{20}$ \cc\ [P\'eroux et al. 2002a], 
because we are considering as DLAs the QALs with N(HI)$>10^{20}$ \cc;
and LLSs are QLAs with N(HI)$>1.6\times 10^{17}$ \cc\ [Tytler 1982])
For a small impact parameter ($r=1$ kpc), the model
satisfies the simultaneous constraints on H(HI) and metal column density of DLAs
only during its early evolution ($t=0.02-0.2$ Gyr). 
At larger impact parameters, these constraints are satisfied 
during a long time span (from $t \sim 0.03$ to $>1$ Gyr).
We did not try to tune the model to exactly match the
observed values of N(HI).
However, only a variation of a factor of two in the masses 
of the models accounts for the whole distribution
of DLAs in the log N(HI)--metallicity plane, as we see from Figure 16,
where we show the results for 
a model with $\nu=3$ Gyr$^{-1}$ and $M_G=5\times 10^8$ \msun,
and another with $\nu=1$ Gyr$^{-1}$ and $M_G=5\times 10^8$ \msun.

\section{Discussion}

\subsection{The abundance ratios}

The presence of dust in DLAs is suggested
by the undersolar [Cr/Zn] and [Fe/Zn] ratios (Pettini et al. 1994, 1999). 
This behaviour is generally attributed
to the fact that Fe and Cr are refractory elements
while Zn is little depleted, as seen in the ISM of the Galaxy.
However, dust depletion is not so clearly indicated by the $\alpha$ elements:
in most of the systems the [S/Si] ratio is close to solar
(but there are more DLAs with suprasolar [S/Si] than with undersolar [S/Si]).
The values above solar could be
explained by taking into account the depletion pattern of the local ISM: 
while Si is depleted, S is nearly undepleted. 
On the other hand, the undersolar values, if real, are difficult to
explain in these context.
They may represent intrinsic values in DLAs, 
where the nucleosynthesis of S could result of contributions 
distinct from those of the local ISM,
or could be explained in terms of differences in the
ionization corrections for S and S.

In DLAs, the [Si/Fe] ratios, even after dust-correction,
reveal the same $\alpha$-element enhancement
observed in metal-poor stars of the Galactic disk.
However, there are systems with nearly solar [Si/Fe] ratios,
even at low metallicities, which is similar to the pattern 
predicted by dwarf galaxy models.
This is consistent with a multi-population scenario for DLAs.
However, the  dwarf galaxy chemodynamical model
accounts for nearly all the DLA data,
while there are a number of systems not explained by the disk model.

The pure chemical evolution models, both for disk systems and dwarf galaxies,
present an apparent discrepancy 
(which is more serious for the dwarf galaxy models)
between the fits to the [Si/Fe] and [N/$\alpha$]i ratios.
In the disk model, 
the [Si/Fe] ratios are better reproduced at larger radii than those
required by the [N/$\alpha$] ratios,
and, in the dwarf galaxy model, by lower values of $\nu$ than those
implied by the [N/$\alpha$] ratios. 
This difference could be related to an underestimated dust correction for Si,
but it does not seem to be the case, 
since the trends [Si/Fe]$_{cor}$ vs. [Fe/H]$_{cor}$
and [S/Zn] vs. [Zn/H] are similar.

Another effect that could contribute to the above discrepancy is that
different samples are used in the comparisons to the models.
The DLA sample with [N/$\alpha$] values is smaller than
the sample with [Si/Fe]$_{cor}$ values.
If a multi-population scenario for DLAs applies,
some galaxy type missing or a few appearances of some rare galaxy type
in the small number DLA sample with [N/$\alpha$] values
could make difficult the comparison between the two samples.
In addition,
the sample with [N/$\alpha$] values has higher redshifts
($z_{\rm median}=3.02$)
than the sample with [Si/Fe]$_{cor}$ values
($z_{\rm median}=2.07$),
then one cannot exclude the possibility that the samples
are drawn from different populations
(different galaxy types or galaxies at different evolutionary stages).

Differently from one-zone models, the chemodynamical model
accounts for nearly all the DLA data.
One single model (the $\nu$ = 3 Gyr$^{-1}$, $M = 10 ^9$ $\msun$ model)
is able to reproduce
[$\alpha$/Fe] and [N/$\alpha$] ratios and HI column densities
of five out of six DLAs,
if we consider a range of radii for the DLA sites
--- from central ($r\sim 1$ kpc) to 
relatively outlying regions ($r\sim 7$ kpc).
Therefore, from the chemical point of view,
the DLAs could represent a single 
type of system (the dwarf spheroids in our case).
On the other hand, direct imaging (Le Brun et al. 1997; Pettini et al. 2000)
and kinematical data (Prochaska et al. 1998)
indicate that at least some DLAs are associated to disks.
It is this sort of non-chemical evidence which makes a case for
a multi-population scenario for DLAs.

We should note that for the system of our sample with the highest [Si/Fe],
the $z_{abs}=3.025$ DLA toward Q0347-38,
Levshakov et al. (2002) report [Si/Fe]$=0.95\pm 0.02$,
which is significantly higher than the value 
[Si/Fe]$=1.17\pm 0.03$ of Prochaska et. (2002b) listed in Table 1.
The reason for this difference is the lower $N({\rm Si}^+)$
measured by Prochaska et. (2002b), and not a higher H(HI),
because Prochaska et. (2002b)
adopt $\log H({\rm HI})=20.626\pm 0.005$ from Levshakov et al. (2002).
Prochaska et. (2002b) suggest that the discrepancy is due to
significantly telluric line-blend in the spectrum obtained with the UVES-VLT
by Prochaska et. (2002b).
For this reason, we have adopted the [Si/H] determination
of Levshakov et al. (2002).
However, if the high [Si/H] is confirmed by further observations, 
it will pose problems
both for the dwarf galaxy chemodynamical model and for the disk model,
because the resulting [Si/Fe]=0.74 will be much higher
than the predictions of any model presented here.
If this is the case, it may represent a system with a 
SFR higher (e.g. $\nu=10$ Gyr$^{-1}$) than that assumed in the
dwarf galaxy chemodynamical model ($\nu=3$ Gyr$^{-1}$).

\subsection{The role of gas flows in DLAs}

The failure of the one-zone model for dwarf galaxies
in reproducing the abundance ratio data
is related to a general inability of the classic wind models,
which lead Matteucci et al. (1997) to propose
a multi-burst model for DLAs,
based on a single-zone chemical evolution model for dwarf irregular,
with a series of parameters describing the burst 
-- burst duration, number of bursts, interval between bursts.
In the chemodynamical model, multiple star bursts appear in a natural way,
and no ad hoc parameters describing them are needed.
The inner ($r \la 4$ kpc) region
exhibits a series of short bursts of star formation, 
which can been seen very clearly in the [Si/Fe] vs. [Fe/H] plot,
and which cover all the values derived for the DLAs.
The connection between the gas flow evolution and the star formation rate
is the reason behind the ability of the chemodynamical model in reproducing
the wide range of abundance ratios seen in DLAs.

The success of the chemodynamical model for dwarf galaxies does not
represent an argument against the disk scenario for DLAs, because we chose
a simple model for the disk evolution, that with a single infall of
primordial gas, in order to discern the effect of gas flows in the evolution
of the system (for the same reason, we studied the classic galactic wind
model for dwarf galaxies).
The disk paradigm for DLAs should be tested by more sophisticated models,
such as the double infall model for the Galaxy (Chiappini et al. 1997).
Moreover, the fact that this model considers a more elaborated form for infall
stress the importance of gas flows in chemical evolution of galaxies.
Kinematical data, revealing the velocity field underlying the DLAs
would be very useful for checking distinct DLA scenarios.
However, as a preliminary conclusion, our results seem to exclude
an outflow scenario for DLAs, since our model resolves outflows,
and no DLA is found in the outflow stage/region.

\subsection{The epoch of galaxy formation}

An important implication of the study of the chemical evolution of
DLAs is to investigate the cosmic progress of galaxy formation.
The models provide chemical clocks to infer the
formation epoch of spheroids using the DLA chemical data as constraints.
We will use as simultaneous constraints on the epoch of \gfor\ $z_{GF}$,
the [N/$\alpha$] vs. [$\alpha$/H] and [Si/Fe] vs. [Fe/H]
data of six DLAs,
which are compared with the predictions of the  chemodynamical models 
with $M_G=10^9$ \msun\ and $\nu=3$ and 1 Gyr$^{-1}$.
The systems can be divided in 2 groups according to the region of the
galaxy reproducing the [N/$\alpha$] ratio:
the inner DLAs, located at $r \sim 1$ kpc
(DLAs toward Q1104-1805, Q2243-6031 and Q0347-38),
and the outer DLAs, in the $r \sim 3 - 7$ kpc region
(DLAs toward Q0000-2620, Q0100+13 and Q1331+17).
The model with  $\nu=3$ Gyr$^{-1}$ best fit the inner DLA data.
The system toward Q0000-2620 is fitted only by the $\nu=1$ Gyr$^{-1}$ model,
but it probably has $M_G=2-3\times 10^9$ \msun, from the constraints on N(HI).
We can see that the inner DLAs
would be within very young galaxies, with ages $\sim 0.1$ Gyr.
Only far from the centre, we find older systems:
the DLAs toward Q0000-2620, Q0100+13 and Q1331+17, 
with ages of 0.1-0.5, 0.1-0.3 and 0.5-1.0 Gyr, respectively,
corresponding to $z_{GF}=$ 3.558-4.411, 2.392-2.574, and 2.075-2.465.
The galaxy formation epoch of a DLA in very young environment
is, of course, only slightly higher than its $z_{abs}$.
Assuming an age of 0.1 Gyr,
the DLAs toward Q0347-38, Q1104-1805, and Q2243-6031
have been formed at $z_{GF}=$ 3.160, 1.711, and 2.414, respectively.
These numbers characterize a typical epoch of formation for these systems
around $z_{GF} \sim 2-3.5$ within a range $z_{GF}=1.7-4.5$. 
Higher values of $z_{GF}$ could be achieved if we consider
the systems with highest values of [Fe/H], supposed to be 
the most chemically evolved, and good candidates to be old systems.
This is the case of the DLAs toward
Q0551-366 ($z_{abs}=1.962$) and Q0201+36 ($z_{abs}=2.463$),
with [Fe/H]$_{cor}=$ -0.013 and -0.191 respectively,
which are in a region of the [Fe/H] vs. [Si/Fe] plane
corresponding to galaxy ages of 1-2 Gyr.
Very early galaxy formation could occur:
$z_{GF}=$ 2.791-4.579 and 3.794-8.147,
for Q0551-366 and Q0201+36, respectively.
One general conclusion is that DLAs constitute a relatively young population
with ages not exceeding 2 Gyr, or even 1 Gyr.

\begin{figure}
\centerline{
\psfig{figure=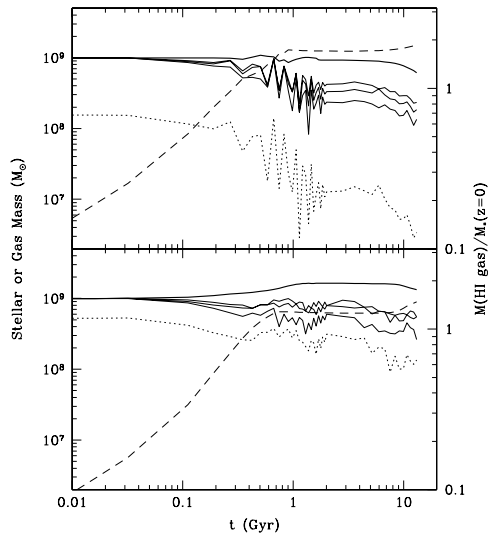,width=8.5cm,angle=0}}
\caption{
Evolution of the baryonic content of the models with $M_G=10^9$ \msun, 
for $\nu=3$ Gyr$^{-1}$ (upper panel) and $\nu=1$ Gyr$^{-1}$ (lower panel).
The thin solid lines give the evolution of the mass in neutral gas
for three temperature limits 
($T \leq 10^4$, $2\times 10^4$ and $3\times 10^4$ K).
The solid line is the evolution of the total amount of gas
(at any temperature).
The dashed line represents the evolution of the stellar mass.
Also shown the evolution of the 
neutral gas fraction with respect to the stellar mass at $z=0$,
assuming that the galaxy has been formed at $z=4.5$ (dotted line).
}
\end{figure}

\subsection{Neutral gas content of DLAs}

While the directly observable baryonic content of galaxies
at the  present epoch is concentrated
in stars, in the past, this must be in the form of gas.
The near coincidence of the
inferred baryonic mass density of DLAs ($\Omega_{DLA}$) at $z \sim 2$ and the
mass density in stars today (Storrie-Lombardi \& Wolfe 2000)
makes the DLAs the prime candidates to the progenitors of modern galaxies.
Since the  DLAs are the dominant reservoir of neutral (HI) gas at $z>0$
available for galaxy formation,
it is of interest to establish whether 
the dwarf galaxy chemodynamical model
predicts enough HI gas to make the DLAs important contributors
to the present day stellar mass of the universe.
In addition, the evolution of the H I mass in our models should be
compared to recent estimates of the evolution of $\Omega_{DLA}$
(Rao \& Turnshek 2000; Storrie-Lombardi \& Wolfe 2000; 
P\'eroux et al. 2001).

Figure 17 shows the \ev\ of the HI gas for the models
with $M_G=10^9$ \msun, and $\nu=3$ Gyr$^{-1}$ and $\nu=1$ Gyr$^{-1}$.
We assume that the neutral gas contains only HI and no H$_2$,
since our calculations do not include molecules.
Our adopted cooling function is valid down to $T=1900$ K,
where the H$_2$ fraction is still unimportant.
Any gas that cools below this temperature
is removed from the fluid according to
our prescriptions for thermal instabilities
(Fria\c ca 1993, Fria\c ca \& Jafelice 1999, Gon\c calves \& Fria\c ca 1999),
and is used in \sfor.
The identification of the HI gas to the neutral gas
is supported by the fact that
in DLAs $N$(H$_2$)\,$\ll N$(H~I) (Petitjean, Srianand \& Ledoux 2002).
We consider as HI gas that with $T \leq2 \times 10^4$ K.
This upper temperature limit
is somewhat higher than the typical $\approx 10^4$ K
of the warm medium in the Galactic disk,
but in the halo warm medium, the temperatures may be considerably
higher than in the disk warm medium (Sembach \& Savage 1996).
In addition, Prochaska et (2002a) derived from the Doppler parameters
of the lines of the $z_{abs}=2.625$ DLA toward GB1759+7539
the upper limit $T < 3\times 10^4$ K.
In order to assess the choice of an upper temperature limit
for the HI gas,
Figure 17 shows the \ev\ of the mass in HI gas,
assuming three upper temperature limits 
($T \leq 10^4$, $2\times 10^4$ and $3\times 10^4$ K),
and also the evolution of the total amount of gas, at any temperature.
All the models are run until $t=13$ Gyr.
We see that, at late times $t>1$ Gyr, the high temperature gas
makes an important contribution to the total gas content of the galaxy.
Most of the hot gas is confined to the outer regions of the galaxy
($r \ga 7$ kpc) and can achieve temperatures higher than $10^6$ K.
In a complete census of metal and mass in gas, this hot gas
should be included in addition  to the HI gas.
Figure 17 also shows the evolution of the stellar mass,
and of the HI gas fraction with respect to the
stellar mass at $z=0$, assuming a galaxy formation epoch $z_{GF}=4.5$.
The HI gas fraction allows us to relate the HI gas in DLAs
to the present day galaxy stellar content.
As we see, in the model with $\nu=1$ Gyr$^{-1}$
the baryonic content is dominated by the total gas at all times,
and the stellar mass becomes larger than the HI gas mass only for $t\ga 10$ Gyr.
By contrast, in the model with  $\nu=3$ Gyr$^{-1}$,
the gas is quickly consumed by star formation,
and the fraction of HI gas is low during most of the evolution of the model.

\begin{figure}
\centerline{
\psfig{figure=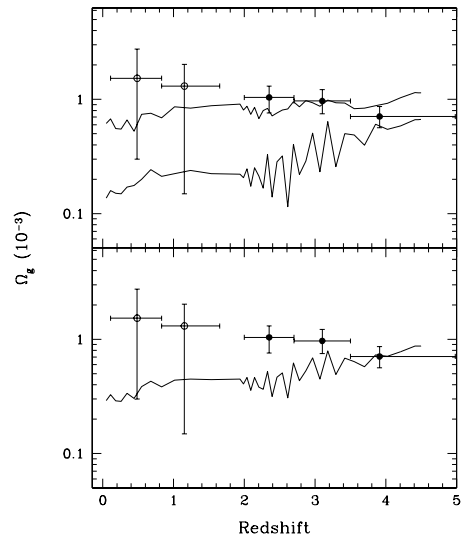,width=8.5cm,angle=0}}
\caption{
Comparison of the observed evolution of the $\Omega_g$ due to the DLAs 
to the predictions of the dwarf galaxy chemodynamical models 
with $M_G=10^9$ \msun, 
assuming that the dwarf galaxy population has been formed at $z=4.5$.
The upper panel shows the predictions of the models
$\nu=3$ Gyr$^{-1}$ (lower curve) and $\nu=1$ Gyr$^{-1}$ (upper curve),
and the lower papel shows the contribution of a 1/2-1/2 mixture
of models with $\nu=1$ and $\nu=3$ Gyr$^{-1}$.
The open circles at low redshift
are the measurements from Rao \& Turnshek (2000),
while the filled circles are the high redshift data from P\'eroux et al. (2001).
Vertical error bars correspond to 1-$\sigma$
uncertainties and the horizontal error bars indicate bin sizes.
}
\end{figure}

Figure 18 shows the evolution of the cosmic HI gas density
$\Omega_g$ (in units of the critical density)
predicted the models
with $M_G=10^9$ \msun, and $\nu=3$ Gyr$^{-1}$ and $\nu=1$ Gyr$^{-1}$,
compared to the observed $\Omega_{DLA}$  evolution.
We combine the HI gas fraction calculated from our models
with the recent estimate of $\Omega_m$ in stars 
in the local Universe by Cole et al. (2001),
$\Omega_*(z=0)=(0.0037\pm0.00056)h_{70}^{-1}$.
We have assumed that the models are typical of the dwarf galaxy population,
and that population was formed at $z_{GF}=4.5$
and has evolved until $z=0$,
giving rise to the dwarf galaxy contribution to $\Omega_*(z=0)$,
i.e. $\Omega_{dwarf}=\Omega_*(M<0.2 M^*)=0.00098h_{70}^{-1}$,
where
$M^*=(1.44\pm0.37)\times 10^{11})h_{70}^{-2}$ \msun\ is the characteristic mass
of the $z=0$ stellar mass function (Cole et al. 2001).
The $\nu=1$ Gyr$^{-1}$ model is potentially a more important contributor
to $\Omega_{DLA}$, due to its larger HI fraction.
However, even in the scenario in which each model 
(with $\nu=3$ Gyr$^{-1}$ and $\nu=1$ Gyr$^{-1}$)
contributes to a half of $\Omega_{dwarf}$, 
the dwarf galaxy population can 
contribute to a half of $\Omega_{DLA}$ at $z>2$, 
and accounts for nearly all $\Omega_{DLA}$ at $z>3.5$,
although at low redshifts its contribution is less important.
Therefore, the dwarf galaxies can be an important contributor to
$\Omega_{DLA}$, specially at high redshifts.

It is interesting to note that 
Boissier, P\'eroux \& Pettini (2003),
using a family of models for disk galaxies,
have found that also the disk galaxies could account 
for about half of $\Omega_{DLA}$,
however, the contribution of the disk population becomes less significant
at high redshift, which is the opposite trend of that we find for the
dwarf galaxies.
This complementarity between the two scenarios, again,
argue for a multi-population scenario for DLAs,
in which dwarf galaxies are more important at higher redshifts
and disks dominate at low redshifts.

\subsection{The high redshift cosmic metal content 
and the relation between DLAs and LBGs}

One possible consequence of dust depletion are selection effects
in optically selected samples.
Boiss\'e et al. (1998) identified a trend of decreasing metallicity 
with increasing N(HI) for DLAs.
They interpreted the lack of systems with large N(HI) and high
metallicities as a selection effect due to dust obscuration:
systems with large N(HI) and high metallicity 
(and consequently high dust content) would obscure the background QSO,
and would not be selected in an optical survey.
This suggestion needs future confirmation,
since, in a QSO radio-select survey of DLAs (Ellison et al. 2001b),
it was found that the bias due to obscuration of QSOs
by the dusty DLAs seems not to be so important. 

The presence of dust depletion and possible selection effects rises the
question of a possible underestimate of the typical metallicity
not only of DLAs but also of other high $z$ systems.
Such could be the case of LBGs.
The success of the chemodynamical model for spheroids 
in reproducing the properties of LBGs (Fria\c ca $\&$ Terlevich, 1999)
gives us confidence in the model prediction that
the LBG metal content is easily underestimated
from a straightforward interpretation of the observations.
The main spectroscopic properties of the 
LBG population at high {\it z} resembles those of star forming 
galaxies (Steidel et al. 1996b), with spectra characterised by a blue 
ultraviolet continuum with moderate dust extinction, strong 
interstellar absorption, P Cygni CIV and NV lines from massive 
stars, and weak Lyman $\alpha$ emission (Lowenthal et al. 1997, 
Pettini et al. 1998, 2000b). 
Although the LBGs are very important objects in the study of the chemical 
enrichment of the early universe because they are the most obvious
star forming regions at high {\it z}, they are not suited 
for chemical abundance studies. Despite the high quality of their spectra,
the large equivalent width of the saturated 
interstellar lines and the high velocity dispersions in the 
ISM of these galaxies make abundance determination in LBGs difficult. 
However, combining observational data and modelling
allows one to infer typical LBG metallicities 
in the range of 0.3 $\Zsun$ to $\Zsun$ (Fria\c ca \& Terlevich 1999).

\begin{figure}
\centerline{
\psfig{figure=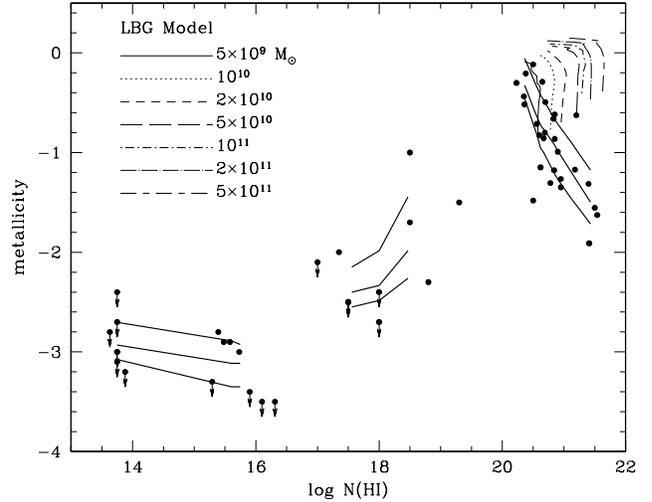,width=8.5cm,angle=0}}
\caption{Metallicity vs. N(HI) data for 
several classes of high {\it z} objects:
LAF clouds, LLSs and DLAs, in order of increasing N(HI).
For DLAs, the metallicity is given by [Si/H]$_{cor}$ 
derived in this paper,
and, for LLS and LAF, it is given
by [O/H], estimated from the observed [C/H] 
(LAF data taken from Lu 1991 and Lu et al. 1998a,
and LLS data from Steidel 1990)
assuming [O/C]= 0.3 (Viegas et al. 1999).
The curves in the right-upper corner are 
the predictions for LBGs from chemodynamical models with several masses,
and the metallicity is [O/H], averaged over a 10 kpc projected radius;
the evolution is shown from 0.1 Gyr (lowest point of the curve) until 1 Gyr.
The thick solid lines are the statistical fits to the data.
}
\end{figure}

Another question is whether DLAs are related to the LBGs.
Mo et al. (1999) proposed that LBGs and DLAs form distinct populations:
the LBGs would be star forming galaxies with low angular momentum and short
collapse time scale, 
while the DLAs would be the initial stages of
evolution of galaxies with high angular momentum and longer
collapse times.

In order to investigate these questions, Figure 19 compares
metallicity vs. N(HI) trends obtained in this work for DLAs
not only to predictions of the chemodynamical model for LBGs
(Fria\c ca \& Terlevich 1999),
but also to data of Ly$\alpha$ forest (LAF) clouds and LLSs. 
As we see from Figure 19, the most metal-rich DLAs could be 
$M_G \la 10^{11}$ \msun\ LBGs in their early evolutionary phases
(the $10^{11}$ \msun\ model is the progenitor of a present-day
$\sim 0.3$ $L^*$ spheroid), but most of DLAs are not explained as LBGs.
The LBGs typically are sub-$L^*$ systems:
the characteristic velocity dispersion of the emission lines,
$\sigma_{em} \approx 70$ km$^{-1}$ (Pettini et al. 1998; Pettini et al. 2001),
is reproduced by $M_G \approx 10^{10}$ \msun\ models
(Fria\c ca \& Terlevich 1999). 
However, it seems that most of the DLA galaxies would be still fainter
than the spectroscopically detected LBGs. 
Good fits to the abundance data are obtained for the 
$\approx 10^9$ \msun\ model.

It is interesting to note that other lines of evidence point toward
to masses not far from $\sim 10^9$ \msun\ as typical for DLAs.
A recent Arecibo blind 21 cm survey (Rosenberg \& Schneider 2003)
has revealed that the number of systems per unit redshift is dominated
by galaxies with HI masses near $10^9$ \msun.
Zheng \& Miralda-Escud\'e (2002) modelled DLAs as spherical isothermal
gaseous halos ionized by the extralactic ionizing background,
and concluded that self-shielding implies a DLA HI mass
of $\approx 1.6\times 10^8$ \msun\ for N(HI)$\approx 3\times 10^{20}$ \cc.

Notice that the normalization of the \sfor\ rate for the LBG models
is $\nu=10$ Gyr$^{-1}$, whereas for the DLA models shown in this paper,
$\nu=3$ or 1 Gyr$^{-1}$. 
The \sfor\ proceeds slowly in DLAs, which would give support to
the scenario for galaxy formation 
in which there is a dichotomy between the star formation
modes for DLAs (long star formation time scale) and LBGs 
(short star formation time scale).
However, the \sfor\ in DLAs does not need to be much
different from that of typical LBGs, since a normalization of the SFR
only a factor $\sim 3$ smaller accounts the
properties of a number of DLAs, thus allowing for a smoother
progression in the evolutionary history of DLAs and LBGs
rather than a sharp dichotomy between the two populations
(see M\o ller et al. 2002).

A complete analysis of the metallicity vs. N(HI) relation for DLAs
should take into account the evolution of metallicity with the redshift,
and the fact that this evolution depends on the column density, 
and, in particular, 
that no evolution with redshift is observed at log N(HI)$>21$
(Boiss\'e et al. 1998, Savaglio 2000).
On the other hand, for systems with low column densities, log N(HI)$<20.8$,
the metallicity does decrease with the redshift (Savaglio 2000).
In any case, the gaps in Figure 19 highlight
domains in column density for which further investigation is required.
There is some suggestion of underestimate of the metal content in DLAs
from the fact that near log(N(HI))$\approx 20.5$, there seems to be
a smooth transition between LBGs and DLAs, whereas at higher
HI column densities, there is a gap between the two classes of objects.
It is possible that a whole class of objects,
intermediate between the observed DLAs and LBGs,
are missing at log(N(HI))$\ga21$.
These could be high metallicity dusty DLA-like objects
or LBG-like objects that are too faint or have  too red $G-{\cal R}$ colours
to be classified as LBGs. 
Also on the low N(HI) side of the DLA distribution it is possible
to hide significant amounts of metal.
The trend of increasing metallicity with decreasing N(HI)
extends down to the lowest values of N(HI) accepted for DLAs
(log(N(HI))$\approx 20$), and then there is a new gap, 
filled with a few LLSs and sub-DLAs. 
This gap suggests that a large number of these objects is missing.
In addition, the ionization corrections for these systems
are very complex, and their chemical abundances easily could be
underestimated.
Proper consideration of the radiation fields
and adequate application of photoionization codes
could lead to metallicity estimates for some LLSs
in the range of 0.1 $\Zsun$ to $\Zsun$
(Viegas \& Fria\c ca 1995).
The ionization corrections should include high ions,
which can make a sizeable contribution to the total metal column density,
specially if we take into account that, 
as said in Section 4.3, these systems
are preferentially located in the outer regions
of the galaxy, where the hot gas concentrates and gives rise
to a local EUV/soft X-ray ionizing radiation field
(Viegas \& Fria\c ca 1995, Viegas et al. 1999).
In this sense, the data points of LLSs in Figure 19
may represent lower limits to the actual metallicities of these systems.
Therefore, these low column density systems might represent
important reservoirs of metals at high redshifts (see P\'eroux et al. 2001).
If we are missing important populations of chemically evolved objects,
the overall metal content in the Universe at high redshift has 
been underestimated, in this way alleviating somewhat the problem
that the density of metal detected at high redshift is
smaller than the integrated total metal production associated
with the star formation activity at high-redshift (Pettini 2000,
Pagel 2000).

\section{Summary and Conclusions}

In this paper we have explored several scenarios for the nature of the DLAs
by means of a study of the metallicity evolution of these systems.
We have analysed the observational data on chemical abundances of DLAs
with robust statistical methods,
and corrected the abundances for dust depletion.
The results of this analysis
have been compared to predictions of several classes of chemical evolution
models describing a variety of scenarios for DLAs:
one-zone dwarf galaxy models, multi-zone disk models,
and chemodynamical models representing dwarf galaxies.
We have found that the chemodynamical model for dwarf galaxies
is more successful than our dwarf galaxy one-zone models and the disk models
in reproducing the properties
--- abundances, abundance ratios, and HI and metal column densities --- of DLAs.
We have then focused on the dwarf galaxy scenario for DLAs, 
as described by the chemodynamical model,
and investigated a number of issues:
the epoch of formation of the host galaxies of DLAs,
the contribution of dwarf galaxies to $\Omega_{DLA}$,
the multi-population (disks and dwarf galaxies) scenario for DLAs,
the relation of DLAs and LBGs,
possible missing populations of high redshift objects,
and ``hidden" reservoirs of metals at high redshifts.
We summarise our results as follows.

1) The $\alpha$-enhancement observed in DLAs seems to be real, even after
dust corrections have been applied to the data. 
In our sample, the observed [Si/Fe]$_{median}=0.32$
is lowered to [Si/Fe]$_{median}=0.16$ by the dust correction,
but remains typically suprasolar,
a pattern similar to that found in  metal-poor stars of the Galactic disk.
However, there are DLAs with nearly solar [Si/Fe] ratio
even at low metallicities,
which could be hosted by dwarf galaxies instead of by disks.

2) The bimodality in the [N/$\alpha$] vs. [$\alpha$/H] distribution of DLAs
pointed out by Prochaska et al. (2002b) is less apparent in our sample.
Instead we find a lower bound envelope
in the [N/$\alpha$] vs. [$\alpha$/H] distribution,
above which the DLAs are concentrated,
and that corresponds to a local specific star formation rate 
$\nu \approx 3$ Gyr$^{-1}$.
This envelope is close to the evolutionary tracks of the
one-zone dwarf galaxy model with $\nu=3$ Gyr$^{-1}$,
and of the disk model at $r=4$ kpc,
but its shape is best reproduced 
by the dwarf galaxy chemodynamical model at $r=1$ kpc.

3) The one-zone models, both for disk systems and dwarf galaxies,
present an apparent discrepancy between fits 
using [Si/Fe] and [N/$\alpha$] ratios.
In the disk model, 
the [Si/Fe] ratios are better reproduced in larger radii than those
required by the [N/$\alpha$] ratios,
and, in the dwarf galaxy model, by lower values of $\nu$ than those
implied by the [N/$\alpha$] ratios. 

4) The chemodynamical model for dwarf galaxies
is more successful than the pure \chev\ models
in reproducing the properties in DLAs.
The connection between the gas flow evolution and the star formation rate
is the reason behind this ability of the chemodynamical model.
The chemodynamical model presents multiple star bursts,
fed by gas flows
which lead the evolutionary tracks of the model 
to cover all the values exhibited by 
the [Si/Fe] vs. [Fe/H] and [N/$\alpha$] vs. [$\alpha$/H] distributions
of the DLAs.
A multi-burst model for DLAs
have already be proposed by Matteucci et al. (1997),
but, in the chemodynamical model, the starbursts appear in a natural way,
with no need of additional parameters describing them.

5) From the study of DLAs with simultaneous measurements of
[N/$\alpha$] and [$\alpha$/Fe] ratios
we have inferred the ages and \for\ epoch  $z_{GF}$ of DLAs.
We have found that $z_{GF}=1.7-4.5$,
and ages ranging from 0.1 Gyr to 1 Gyr.
Higher values of $z_{GF}$ could be achieved if we consider
the systems with highest values of [Fe/H], supposed to be 
the most chemically evolved, and good candidates to be old systems.
In this case, we find $z_{GF} \sim3 - \sim 8$.
One general conclusion is that DLAs constitute a relatively young population
with ages not exceeding 2 Gyr, or even 1 Gyr.

6) The dwarf galaxies can
contribute to about half of $\Omega_{DLA}$ at $z>2$, and even more, at $z>3.5$,
although at low redshifts their contribution is less important.
A multi-population scenario for DLAs seems to favoured,
in which the DLA population is dominated 
by dwarf galaxies at high redshifts and by disks at lower redshifts.

7) The results of the chemodynamical model calls for a smoother
progression in the evolutionary history of DLAs and LBGs
rather than a sharp dichotomy between the two populations.
LBGs and DLAs could constitute a sequence of increasing star formation
rate, with the LBGs being the systems 
with typically short star formation time scales ($\sim10^8$ yr),
and the DLAs corresponding to the low $\nu$ ($\approx 1-3$ Gyr$^{-1}$)
end of the sequence.
We also arise the possibility that we could be missing a whole population of
objects at high hydrogen density columns with metallicities intermediate
between those of DLAs and LBGs.

8) It is possible that relying only on the observations of DLAs
could lead to an underestimate of
the metal content of the high redshift Universe,
because we are not properly including in the metal census
both the above mentioned high column density population
and the low column density population constituted by LLSs and sub-DLAs.

\section*{Acknowledgments}

We thank Timothy Beers for making available the {\it LOWESS} and {\it
MSTAT} routines to us and for the help with the robust statistical methods. 
We thank the anonymous referee for the detailed comments which improved
this work.
G.A.L. and A.C.S.F. acknowledge financial support from
the Brazilian agencies CNPq, 
FAPESP (proj. 00/10972-0) and CNPq/PRONEX.

\bsp

\end{document}